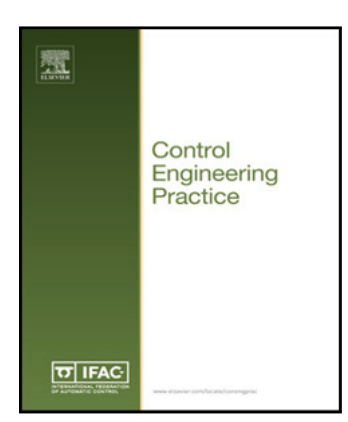

# Model reference-based control with guaranteed predefined performance for uncertain strict-feedback systems

Mehdi Heydari Shahna [a],*, Jukka-Pekka Humaloja [b], Jouni Mattila [a]

[a] Faculty of Engineering and Natural Sciences, Tampere University, Tampere, 33720, Finland
[b] Faculty of Electrical and Computer Engineering, Technical University of Crete, Chania, Greece

## ARTICLE INFO

## ABSTRACT

A wide range of practical applications in robotics and automation can be modeled in a class of uncertain and nonlinear strict-feedback (SF) systems. In SF systems, the hierarchical influence of control inputs on state dynamics renders each level dependent on preceding control actions. However, designing a robust, performance-guaranteed controller is challenging in real-world applications due to the complexities introduced by time- and state-varying uncertainties and the difficulty in computing analytic derivatives in SF systems. To address this challenge, this study introduces a novel model reference-based control (MRBC) framework that applies locally to each subsystem (SS) of SF systems, to ensure output tracking performance within the specified transient and steady-state response criteria. This framework includes (1) novel homogeneous adaptive estimators (HAEs) designed to match the uncertain nonlinear SF system to an ideal reference model, enabling easier analysis and control design at the SS level, and (2) model-based homogeneous adaptive controllers enhanced by logarithmic barrier Lyapunov functions (HAC-BLFs), intended to control the reference model provided by HAEs in each SS, while ensuring the prescribed tracking responses under control amplitude saturation. The inherently robust MRBC achieves uniformly exponential stability using a generic stability connector term, which addresses dynamic interactions between the adjacent SSs. The parameter sensitivities of HAEs and HAC-BLFs in the MRBC framework are analyzed, focusing on the system's robustness and responsiveness. The proposed MRBC framework is experimentally validated on an electromechanical linear actuator system with an uncertain SF form, by comparison with two high-performance adaptive control strategies under loading disturbance forces challenging 0–95% of its capacity.

## 1. Introduction

Strict-feedback (SF) form garnered significant focus from 1990s in the realm of recursive control strategies (Li, Li, & Tong, 2022; Wang & Zhu, 2015) as they encompass a wide range of practical applications, including robot manipulators (Doctolero & Westwick, 2023), electromechanical actuators (Zhang & Yu, 2024), DC-DC buck converters (Zhang, Wang, Li, Wu, & Qian, 2014), electropneumatic systems (Smaoui, Brun, & Thomasset, 2006), wheeled systems (Mathieu & Hedrick, 2011), and stimulated quadriceps muscles (Schauer et al., 2005). In SF systems, the hierarchical influence of control inputs on state dynamics renders each level dependent on preceding control actions (Yan et al., 2023). Therefore, they require control approaches to counteract the potential for finite escape instabilities inherent in an open loop. Generally, for a system dynamics expressed as $x = f(x) + g(x)u$, the functional term $g(x)$ multiplied by the control input is the control gain function, the sign of which is referred to as the control direction (Zhang & Yang, 2019). In certain SF systems, the control gain is predetermined (Koivumäki, Humaloja, Paunonen, Zhu, & Mattila, 2022; Liu et al., 2021), or its sign is known (Xie, Jing, Zhang, & Dimirovski, 2024; Zhang & Yang, 2016). However, gaining a prior understanding of the control direction under broader conditions and uncertain applications seems impractical (Lozano & Brogliato, 1992). To address unknown control directions in SF systems, Zhang and Yang (2019) introduced a group of error transformation functions. In addition, Zhang, Deng, and An (2024) assumed that the modeling term $f(x)$ is accurate and fully known in the SF structure, or partially known, as assumed in Koivumäki et al. (2022). Hence, Zhang and Yang (2016) proposed a group of new feedback mechanisms to compensate for the unknown system dynamics. In addition to these conservative assumptions in designing control for SF systems, significant computational complexities, referred to as the "explosion of complexity", have been noted (Ding & Zhang, 2021; Liu et al., 2021; Zhang & Yang, 2019),

* Corresponding author.
*E-mail address:* mehdi.heydarishahna@tuni.fi (M.H. Shahna).

## Nomenclature

### Abbreviations and unit symbols

| | |
|---|---|
| SF | Strict-feedback |
| SS | Sub(system) |
| MRBC | Model reference-based control |
| HAE | Homogeneous adaptive estimator |
| HAC | Homogeneous adaptive controller |
| BLF | Barrier Lyapunov function |
| EMLA | Electromechanical linear actuator |
| FF | Feedforward |
| PPC | Prescribed performance control |
| PMSM | Permanent-magnet synchronous motor |
| AATC | Adaptive asymptotic tracking control |
| AFTC | Adaptive fault-tolerant control |
| DoF | Degree-of-freedom |
| MIMO | Multi-input multi-output |
| m | Meter |
| N | Newton |
| N/$\mu$m | Newton per micrometer |
| N m | Newton-meter |
| N s/m | Newton-second per meter |
| kg m$^2$ | Kilogram square meters |
| m/s | Meters per second |

### Parameters in a SF system

| | |
|---|---|
| $f(x)$ | General system term |
| $x$ | General system state |
| $g(x)$ | General control gain function |
| $u$ | Control input |
| $i$ | Corresponds to the number of SSs |
| $n$ | Order of a SF system ($i = 1, \ldots, n$) |
| $x_i$ | $i$th system state |
| $\boldsymbol{x}_i$ | Vector of system states from the 1st to the $i$th SS |
| $\mathbb{R}^i$ | i-dimensional real coordinate space |
| $f_i$ | Time- and state-variant modeling function of a SF system |
| $\Gamma_i$ | Time-variant External disturbance of a SF system |
| $d_i$ | Time and state-variant triangular uncertainty of a SF system |
| $g_i$ | Time and state-variant functional control gain of a SF system |

### Parameters in HAEs

| | |
|---|---|
| $\hat{x}_i$ | New estimated state of the reference model |
| $f_i^*$ | New estimated modeling term of the reference model |
| $e_i$ | Error between new states in the estimated reference model and actual states in the SF system |
| $\tilde{\Psi}_i$ | Adaptive law in HAEs |
| $\xi_i$ | Design parameter of HAEs |
| $\lambda_i$ | Design parameter of HAEs |
| $\beta_i$ | Design parameter of HAEs |
| $d_i^*$ | Time and state-variant uncertainty in $e_i$ |
| $s_i$ | Stability connector between SS$_i$ and SS$_{i+1}$ |
| $V_i$ | Quadratic function for $i$th HAE |
| $V_{ob}$ | Quadratic function for HAEs |
| $\boldsymbol{e}_{ob}$ | Matching-error vector in HAEs |
| $\ell_{ob}$ | Positive Offset in $\boldsymbol{e}_{ob}$ |
| $\rho_{ob}$ | Positive parameter depending on parameters of HAEs |
| $\zeta_i$ | Positive parameter |
| $g_0(\tau_0)$ | Stability region, centered at zero, of the HAEs |

### Parameters in HAC-BLFs

| | |
|---|---|
| $u_{min}$ | Lower bound of $u$ |
| $u_{max}$ | Upper bound of $u$ |
| $\alpha(.)$ | Amplitude saturation function of $u$ |
| $\alpha_1$ | Functional gain coefficient of $\alpha(.)$ |
| $\alpha_2$ | Functional offset of $\alpha(.)$ |
| $\Delta_u$ | Intensity of saturation in $u$ compared to $\alpha(u)$ |
| $\alpha_{max}$ | Absolute value of $\Delta_u$ |
| $\bar{e}_i$ | Tracking error in HAC-BLFs |
| $x_{id}$ | Reference trajectory in $i$th HAC-BLF |
| $\bar{f}_i$ | FF compensation in $i$th HAC-BLF |
| $\gamma_i$ | Design parameter of HAC-BLFs |
| $\epsilon_i$ | Design parameter of HAC-BLFs |
| $\tilde{\theta}_i$ | Adaptive law in HAC-BLFs |
| $Q_i$ | Positive notation in HAC-BLFs |
| $\kappa_i$ | Design parameter in HAC-BLFs |
| $o_i$ | Tracking performance threshold in HAC-BLFs |
| $o_i^{shoot}$ | Positive parameter for designing $o_i$ |
| $o_i^{bound}$ | Positive parameter for designing $o_i$ |
| $o_i^*$ | Positive parameter for designing $o_i$ |
| $t_0$ | Initial time |
| $\bar{s}_i$ | Stability connector between SS$_i$ and SS$_{i+1}$ in HAC-BLFs |
| $\bar{V}_{cont}$ | Quadratic function for HAC-BLFs |
| $\bar{V}_i$ | Quadratic function for $i$th HAC-BLF |
| $\bar{\rho}_{cont}$ | Positive parameter depending on parameters of HAC-BLFs |
| $\bar{\ell}$ | Positive Offset in $\bar{e}_i$ |
| $\nu_i$ | Positive parameter |
| $g_1(\tau_1)$ | Stability region, centered at zero, of the HAC-BLFs |

### Parameters in MRBC

| | |
|---|---|
| $\bar{V}_{all}$ | Quadratic function for the MRBC framework |
| $\ell_{all}$ | Positive Offset in the stability proof of the MRBC |
| $g_2(\tau_2)$ | Stability region, centered at zero, of the MRBC |
| $\rho_{all}$ | Positive parameter depending on parameters in the MRBC |

### Parameters in a PMSM-driven EMLA

| | |
|---|---|
| $\upsilon_L$ | Linear velocity state in m/s |
| $x_L$ | Linear position state in m |
| $\tau_m$ | Motor torque in N m |
| $I_{eq}$ | Equivalent inertia in kg m$^2$ |
| $B_{eq}$ | Mechanical (viscous) damping in $\frac{\text{N s}}{\text{m}}$ |
| $K_{eq}$ | Spring effects in $\frac{\text{N}}{\mu\text{m}}$ |
| $f_{eq}$ | Load coefficient of the actuator system |
| $F_L$ | Load disturbance in $N$ |

stemming from the excessive growth of analytic derivatives in control designs due to the inherent recursiveness of these systems. To address this, Koivumäki et al. (2022) proposed a specific feedforward (FF) compensation term for each subsystem (SS) of a class of SF systems, based on the system's inverse dynamics. Similarly, command filters were employed in Shen and Shi (2015), to overcome the computational complexity and control singularity caused by the repeated derivation of virtual control functions in backstepping control applied to SF systems. Furthermore, most practical systems typically function within various input $u(t)$ and output $x(t)$ constraints, which may be due to physical limitations, performance requirements, or security considerations (Gaagai & Horn, 2023; Shahna & Abedi, 2021). For example, system input amplitude saturation is common in SF applications and can significantly degrade performance, potentially leading to system instability. Hence, Min, Xu, Zhang, and Ma (2018) proposed a state observer combined with a backstepping recursive approach to ensure the global boundedness of the closed-loop system with high probability. While recent successes have been reported, the challenges related to different types of uncertainties, including external disturbances, modeling and parametric uncertainties, and unknown control gains, affecting tracking control performance, remain inadequately addressed. Some novel control strategies, such as prescribed performance control (PPC), were recently explored in Bikas and Rovithakis (2018) and Theodorakopoulos and Rovithakis (2015) to enhance both transient and steady-state performances while decreasing the computational burden of the control strategy. However, these schemes required the control direction to be known in advance. This problem was addressed in Xie et al. (2024) through the introduction of smooth orientation functions. However, this study overlooked the control input peaking and amplitude saturation phenomenon.

Inspired by the aforementioned insights and building upon (Koivumäki et al., 2022; Nguyen, 2018; Shakarami, Esfandiari, Suratgar, & Talebi, 2020; Zhang & Yang, 2019), this paper proposes a model reference-based control (MRBC) framework for a class of uncertain SF systems. This framework has two functional stages:

(1) Homogeneous adaptive estimators (HAEs): As specific FF compensation terms, they are designed to transform an uncertain nonlinear SF system into a certain reference model. The modeling terms of the reference system at the SS levels are adaptively estimated, while the reference system states are matched to the uncertain nonlinear SF system states of equivalent order. This reference model provides the foundation for the next stage of the MRBC framework.

(2) Homogeneous adaptive controllers enhanced by logarithmic barrier Lyapunov functions (HAC-BLFs): As FF compensation terms for the certain reference model obtained in the first stage, they are designed to actively engage based on the system's inverse dynamics, generating the control input signal $u(t)$ from the reference system states and desired output, thereby enforcing the entire system to adhere to the prescribed tracking control responses at the SS levels.

The schematic block diagram of the overall system in a standard closed-loop configuration, employing the proposed MRBC framework, is illustrated in Fig. 1.

To provide a clearer understanding of how the proposed MRBC framework integrates into a practical system, its key components are summarized as follows.

**Inputs**: the MRBC receives three types of inputs: (1) measured system states obtained from the sensors, (2) reference or desired trajectories, and (3) predefined performance constraints including limits on overshoot, steady-state error, and the control input itself.

**Outputs**: the sole output of the MRBC is the computed control signal that drives the system.

**Design parameters**: the design involves a total of $6n$ tuning parameters, where $n$ is the system order. Specifically, $3n$ parameters are associated with the BLF-HACs, and another $3n$ are used for configuring the HAEs. Tuning parameter sensitivity will be discussed later.

The comparative contributions are as follows:

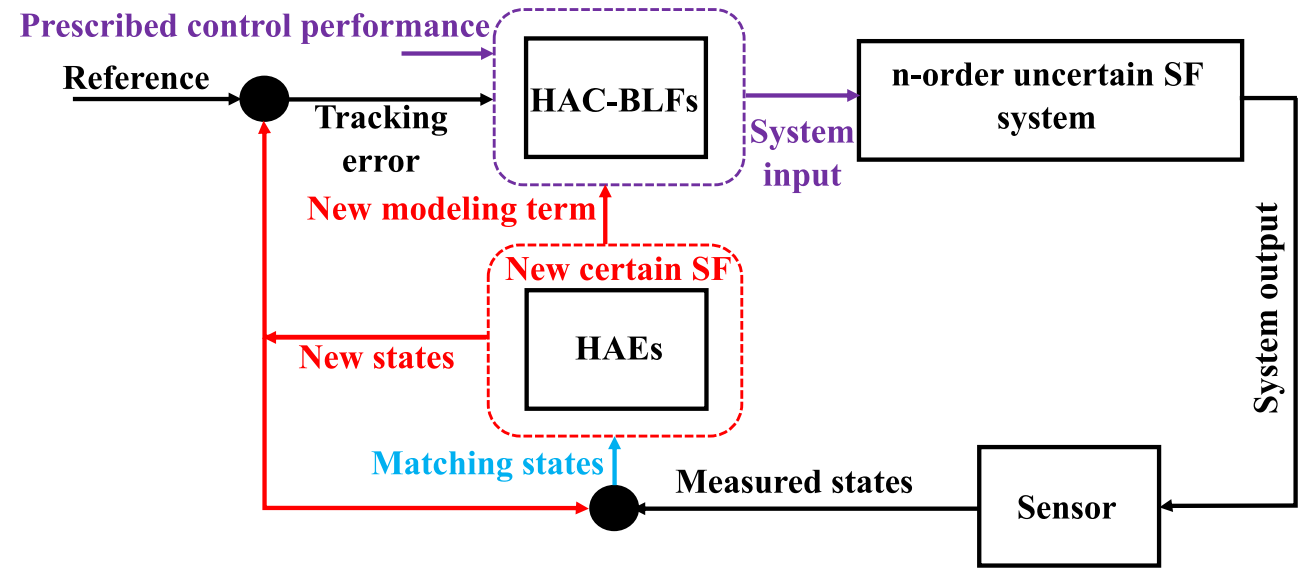

**Fig. 1.** MRBC framework for an $n$-order SF system.

(1) This paper designs a novel performance-guaranteed robust MRBC framework for a comprehensive class of uncertain and nonlinear SF systems by addressing an unknown control gain function and its sign (contrary to Bikas & Rovithakis, 2018; Koivumäki et al., 2022; Liu et al., 2021; Theodorakopoulos & Rovithakis, 2015; Xie et al., 2024; Zhang et al., 2024; Zhang & Yang, 2016), time-varying disturbances, and state-dependent modeling uncertainties (contrary to Koivumäki et al., 2022; Liu et al., 2021; Zhang et al., 2024).

(2) To transform the comprehensive uncertain SF system into a more user-friendly format for easier analysis and control design, this paper proposes novel HAEs. In this way, the modeling terms of the reference system at the SS levels are adaptively estimated, while the reference system states are matched to the uncertain nonlinear SF system states of equivalent order. This reference model provides the foundation for further developments.

(3) Unlike Zhang and Yang (2019), which used mathematical manipulations due to the transformation of asymmetric constraints into symmetric forms of predefined tracking performance, HAC-BLFs guarantee tracking performance within the user-defined transient and steady-state performance by directly constraining the error symmetry form.

(4) Built on Koivumäki et al. (2022) to address the challenges of computing analytic derivatives of virtual controls used in each $i$th subsystem ($SS_i$), where $i = 1, \dots, n$, of an $n$-order SF system, this study eliminates the need for using analytic derivatives of virtual control (contrary to Liu et al. (2021)). This is achieved by designing an MRBC framework based on the system's inverse dynamics to generate the control input from the states and desired output of the system. This paper also extends the stability connector concept proposed in Koivumäki et al. (2022), where the subsequent SS compensates each SS with a "stability-preventing" connector with a corresponding "stabilizing" connector to achieve asymptotic stability. By employing the MRBC framework, the tracking error of the SF system uniformly and exponentially converges to a stable region, whose radius depends on the intensity of the uncertainties. Built on Shakarami et al. (2020), the parameter sensitivities of the MRBC framework are analyzed, with a focus on the trade-off between system responsiveness and uncertainty rejection.

The rest of this paper is organized as follows. Section 2 designs HAEs. Section 3 formulates HAC-BLFs. Section 4 analyzes the control stability. Section 5 validates the proposed framework on a practical SF system, an electromechanical linear actuator (EMLA) system, by comparison with two state-of-the-art high-performance adaptive control strategies, despite the presence of load disturbances. Finally, Section 6 concludes this paper.

## 2. Designing HAEs for an uncertain SF system

### 2.1. Model problem

Let us define $x_i \in \mathbb{R}$, where $i = 1, \dots, n$, as the system states and $\boldsymbol{x}_i = \left[x_1, \dots, x_i\right]^T \in \mathbb{R}^i$, $\boldsymbol{x}_n = \boldsymbol{x} \in \mathbb{R}^n$, and $u \in \mathbb{R}$ as the system input.

As used in Zhang et al. (2024), an ideal $n$-order SF system, especially suitable for model-based control approaches, can be represented as follows:

$$\begin{cases} \dot{x}_i(t) = x_{i+1}(t) + f_i\left(\boldsymbol{x}_i, t\right) \\ \dot{x}_n(t) = u(t) + f_n\left(\boldsymbol{x}_n, t\right) \end{cases} \tag{1}$$

where the time- and state-variant functions $f_i : \mathbb{R}^i \times \mathbb{R}_+ \to \mathbb{R}$ are known as modeling terms with a triangular structure (Pan & Basar, 1998). However, the system representation (1) for many control applications is impractical. Based on Zheng and Yang (2019), this paper addresses a class of model-based SF system representations by considering different systematic uncertainties:

$$\begin{cases} \dot{x}_i = g_i\left(\boldsymbol{x}_i, t\right) x_{i+1} + f_i\left(\boldsymbol{x}_i, t\right) + d_i\left(\boldsymbol{x}_i, t\right) + \Gamma_i(t) \\ \dot{x}_n = g_n\left(\boldsymbol{x}_n, t\right) u + f_n\left(\boldsymbol{x}_n, t\right) + d_n\left(\boldsymbol{x}_n, t\right) + \Gamma_n(t) \end{cases} \tag{2}$$

where the time- and state-variant triangular function $f_i : \mathbb{R}^n \times \mathbb{R}_+ \to \mathbb{R}$ represents a modeling term, if available, which is not strictly required and may vary depending on the specific application. The time-variant function $\Gamma_i : \mathbb{R}_+ \to \mathbb{R}$ is unknown external disturbance. The time- and state-variant function $d_i : \mathbb{R}^i \times \mathbb{R}_+ \to \mathbb{R}$ is unknown triangular uncertainty and local Lipschitz. The function $g_i\left(\boldsymbol{x}_i, t\right) : \mathbb{R}^i \times \mathbb{R}_+ \to \mathbb{R}$ is an unknown, nonzero, time- and state-varying functional control gain. The model-based uncertain SF system presented in Eq. (2) provides a comprehensive representation of many dynamical systems. However, designing an efficient controller for this $n$-order model dynamics is challenging owing to the unknown control gain function, time-varying disturbances, and state-dependent modeling uncertainties.

### 2.2. Transformation of an uncertain into a certain SF system model

As the first aim, this paper transforms the uncertain SF system (2) into the ideal format (1), facilitating subsequent analysis and controller design. This transformation provides the foundation for designing the MRBC framework. For brevity, the notations $t$ and $\boldsymbol{x}_i$ are frequently omitted from the equations. A reference model, based on the ideal model in (1), is defined as follows:

$$\begin{cases} \dot{\hat{x}}_i = \hat{x}_{i+1} + f_i^* \\ \dot{\hat{x}}_n = u + f_n^* \end{cases} \tag{3}$$

where $f_i^* : \mathbb{R}^i \times \mathbb{R}_+ \to \mathbb{R}$ are new estimated modeling terms of the reference system at the SS levels and based on uncertainties and disturbance effects. They are defined based on adaptation errors $\tilde{\Psi}_i$ and estimation errors $e_i = x_i - \hat{x}_i$:

$$f_i^* = f_i + \xi_i \tilde{\Psi}_i e_i + \frac{1}{2}\lambda_i e_i + e_{i-1} \tag{4}$$

where $e_0 = 0$ and $\xi_i$ are positive constants. As mentioned, $f_i$ is not strictly required and may vary depending on the specific application. Homogenous terms $\xi_i \tilde{\Psi}_i e_i + \frac{1}{2}\lambda_i e_i + e_{i-1}$ play estimator roles to match the reference system states to the uncertain nonlinear SF system states (2) with equivalent order. Adaptive laws $\tilde{\Psi}_i$ for each $SS_i$ are defined as:

$$\dot{\tilde{\Psi}}_i = -\beta_i \tilde{\Psi}_i + \xi_i e_i^2 \tag{5}$$

where $\beta_i$ is a positive constant. Now, from the definition of $e_i$ as well as (3) and (4):

$$\begin{aligned} \dot{e}_i &= e_{i+1} + d_i^* + \Gamma_i - \xi_i \tilde{\Psi}_i e_i - \frac{1}{2}\lambda_i e_i - e_{i-1} \\ \dot{e}_n &= d_n^* + \Gamma_n - \xi_n \tilde{\Psi}_n e_n - \frac{1}{2}\lambda_n e_n - e_{n-1} \end{aligned} \tag{6}$$

where

$$d_i^*(\boldsymbol{x}_{i+1}, t) = d_i(\boldsymbol{x}_i, t) + (g_i\left(\boldsymbol{x}_i, t\right) - 1)x_{i+1} \tag{7}$$

**Remark 2.1.** As can be seen from the tracking error model (6), the term $f_i$ cancels out because it is assumed to be a known modeling term. Therefore, the presence or absence of this term does not affect the control design in this paper. Indeed, the presence of $f_i$ can accelerate tracking convergence and reduce control effort. However, the objective of this paper is not to compare model-based and model-free control approaches. Instead, it aims to design a robust and performance-guaranteed controller for a comprehensive class of uncertain SF systems, regardless of whether $f_i$ exists or not. In addition, in some real-world applications, $g_i(\boldsymbol{x}_i, t)$ is known and often takes a constant value, typically $g_i(\boldsymbol{x}_i, t) = 1$. In the mentioned applications $d_i^*(\boldsymbol{x}_i, t) = d_i(\boldsymbol{x}_i, t)$; otherwise, based on Shahna, Bahari and Mattila (2024), $d_i^*(\boldsymbol{x}_{i+1}, t)$ will be non-triangular structures of uncertainties.

**Remark 2.2.** It is commonly assumed that (1) the upper limit of disturbances $|\Gamma_i|$ is bounded and already known Liu and Huang (2006); (2) the partial knowledge of the $d_i(x_i, t)$ function or its corresponding bounding functions are predefined (Huang & Liu, 2019); (3) the bound of $g_i$ is known Oliveira, Peixoto, and Hsu (2015), or at least $g_i(\cdot)$, where $i = 1, \ldots, n$, must follow a structure where it consists of an unknown constant multiplied by a known function (Huang, Wang, Wen, & Zhou, 2018). Apart from the following assumption, this work requires no additional information about the system's nonlinearities.

**Assumption 2.1.** Following the adaptive and robust control literature (see Huang & Liu, 2019; Huang et al., 2018; Liu & Huang, 2006; Xudong, 1999; Zhang & Yang, 2019), the functional gain $|g_i(\boldsymbol{x}_i, t)|$, the modeling uncertainty $|d_i(\boldsymbol{x}_i, t)|$, and external disturbances $|\Gamma_i(t)|$ are assumed to be bounded above by unknown positive constants. This means that for all values of $x_i$ in its domain and for all times $t$, $|g_i(\boldsymbol{x}_i, t)|$, $\left|d_i\left(\boldsymbol{x}_i, t\right)\right|$, and $|\Gamma_i(t)|$ can always be bounded from above by unknown positive values and not grow infinitely large. This assumption is significant, as excessively large uncertainties would require extremely large or even infinite control actions (e.g., torque) to ensure robustness and maintain system stability—actions that are physically infeasible, could potentially damage the actuators, and in practice would typically trigger an emergency shutdown to protect the system. To protect the actuators and ensure practical implementation, Section 3 demonstrates that the control input signal $u$ and the system states remain constrained within defined and bounded values.

### 2.3. Stability of HAEs

Motivated by a key concept in the stability connector proposed in Koivumäki et al. (2022) and the virtual power flow proposed in Zhu (2010), for system (2) with reference system (3) and HAEs provided in (4), new stability connectors are defined as

$$s_i = e_i e_{i+1} \tag{8}$$

This term is defined to capture the triangular dynamic interactions between adjacent subsystems. This auxiliary function is then employed in the convergence analysis of Theorem 2.1 (see Appendix A), demonstrating that the destabilizing interconnections effectively cancel out and that the entire system achieves exponential stability. The error variables $e_i := x_i - \hat{x}_i$ are the differences between the original system signals and the transformed variable. Hence, let us define a quadratic function as

$$V_{\text{ob}}(t) = \sum_{i=1}^{n} \frac{1}{2}\left[e_i^2(t) + \tilde{\Psi}_i^2(t)\right] \tag{9}$$

Assume that there are any positive parameters $\rho_{ob}$, depending on the design parameters of the HAEs as $\rho_{ob} = \min[\min_{i=1}^{n}\left\{\lambda_i - \delta_i - \zeta_i,\ 2\beta_i\right\}]$, and $\ell_{ob}$, depending on the intensity of the uncertainties $d_i^*$ and $\Gamma_i$ as $\ell_{ob} = \sum_{i=1}^{n} \frac{1}{2\delta_i} d_i^{*2} + \frac{1}{2\zeta_i}\Gamma_i^2$. The following theorem is now stated.

**Theorem 2.1.** *Based on (9), the use of (4)–(6) from any initial time $t_0$ guarantees that the matching-error vector $\boldsymbol{e}_{ob} = \left[e_1, \ldots, e_n\right]^T$ converges*

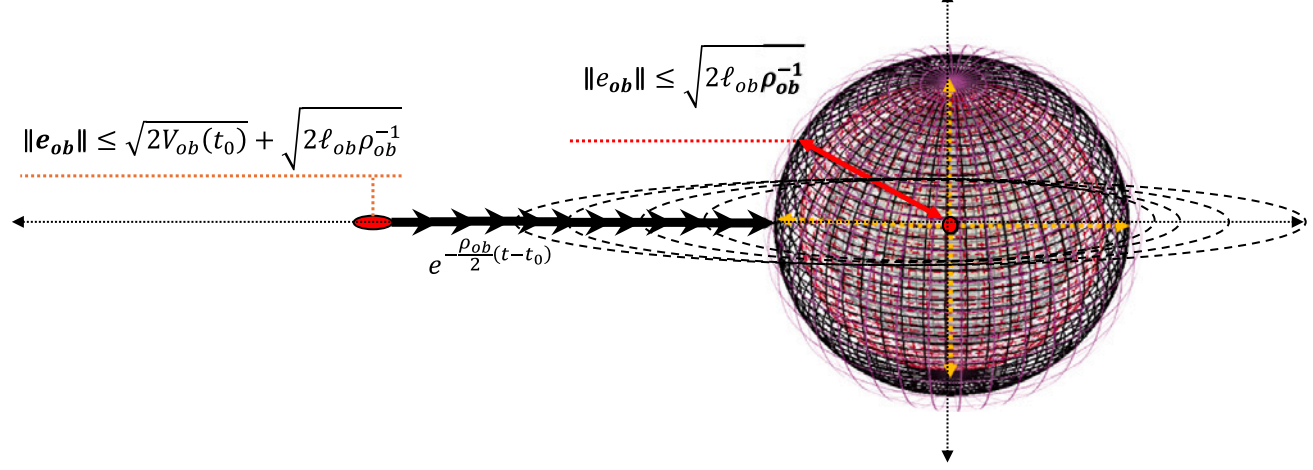

**Fig. 2.** Stable region for HAEs.

**Table 1**
Parameter setting of HAEs, in terms of responsiveness and robustness.

| HAEs' design parameters | Affected parameters | Increase the parameter | Decrease the parameter |
|---|---|---|---|
| $\xi_i$ | None | Insensitive | Insensitive |
| $\min_{i=1}^{n}\{\lambda_i - \delta_i - \zeta_i,\ 2\beta_i\}$ | $\rho_{ob}$ in (42) | Higher response | More robust |

*exponentially with rate $\frac{\rho_{ob}}{2}$ into a stability region of radius $\sqrt{2\ell_{ob}\rho_{ob}^{-1}}$ centered at zero. In other words,*

$$\|e_{ob}\| \le \sqrt{2V_{ob}(t_0)}e^{-\frac{\rho_{ob}}{2}(t-t_0)} + \sqrt{2\ell_{ob}\rho_{ob}^{-1}} \tag{10}$$

**Proof.** See Appendix A.

### 2.4. Parameter sensitivity analysis

Derived from Corless and Leitmann (1993), Eq. (10) expresses that $\|e_{ob}\|$ converges uniformly and exponentially with the decay rate $\frac{\rho_{ob}}{2}$ within a stable region of radius $\sqrt{2\ell_{ob}\rho_{ob}^{-1}}$. Theorem 2.1. is graphically represented in Fig. 2.

As discussed in Mousavi and Guay (2024), a trade-off exists between real-time responsiveness and the robustness capability of the system. This trade-off can be investigated as follows:

(1) Real-time responsiveness: includes the distance of the initial error from the stable region, and the convergence rate into the stable region.

(2) Robustness capability: includes uncertainty-rejecting capability, signifying the stable region radius.

The HAEs provided in (4) for each $SS_i$ require three positive parameters to be selected: $\xi_i$, $\beta_i$, and $\lambda_i$. In Fig. 2, three positive parameters are used to define the stable region–$V_{ob}(t_0)$, $\ell_{ob}$, and $\rho_{ob}$–which are directly affected by the parameters of HAEs, which are shown in Table 1, addressing the relevant equations. For further information, see Appendix A.

For the control parameter $\xi_i$, being positive is sufficient for system stability, and varying it does not affect the control performance. Although the arbitrary parameters $\zeta_i$ and $\delta_i$ are not design parameters of HAEs and their increase reduces the radius of the stable region, $\lambda_i$ must also be increased to ensure that $\rho_{ob} > 0$. Thus, depending on the applications, demands, and intensity of systematic uncertainties, the trade-off between real-time responsiveness and robustness capability, based on Table 1, should be considered.

## 3. Designing HAC-BLFs for tracking control

### 3.1. Control input amplitude saturation

After discussing the reference model transformation in Section 2, as indicated in Assumption 2.1, a constraint must be applied to the system input signal $u(t)$. In this regard, an amplitude saturation function $\alpha(u(t))$ is employed, providing $u_{min} \le u \le u_{max}$, where $u_{min}$ and $u_{max}$ are defined as the lower and upper bounds, respectively. The saturation function in a linear form for subsequent mathematical analysis can be defined as $\alpha(u(t)) = \alpha_1 u(t) + \alpha_2$ as Shahna and Mattila (2024), where

$$\alpha_1 = \begin{cases} \frac{1}{|u|+1}, & u \ge u_{max} \text{ or } u \le u_{min} \\ 1 & u_{min} \le u \le u_{max} \end{cases}$$
$$\alpha_2 = \begin{cases} u_{max} - \frac{u}{|u|+1}, & u \ge u_{max} \\ 0 & u_{min} \le u \le u_{max} \\ u_{min} - \frac{u}{|u|+1} & u \le u_{min} \end{cases} \tag{11}$$

$\Delta_u = \alpha(u) - u = (\alpha_1 - 1)u + \alpha_2$ is defined to represent the intensity of saturation of the control input compared to the constraint one. Employing the control saturation function $\alpha(.)$, from (3):

$$\begin{cases} \dot{\hat{x}}_i = \hat{x}_{i+1} + f_i^* \\ \dot{\hat{x}}_n = \alpha(u) + f_n^* \end{cases} \tag{12}$$

**Assumption 3.1.** Let us assume that the intensity of control input saturation $\Delta_u$, defined as the difference between the imposed constraint and the actual value of the control signal, is bounded and does not grow unbounded. This assumption is reasonable, as in real-world applications, there are physical capacity constraints on generating control input signals. Hence, an unknown positive parameter is defined as $\alpha_{max} = |\Delta_u|$.

### 3.2. Inverse dynamic control

Tracking errors are defined as $\bar{e}_i(t) = \hat{x}_i(t) - x_{id}(t)$, where $x_{1d}$ represents the bounded and differentiable reference trajectories for $SS_1$ within the domain $-\eta_1 \le x_{1d} \le \eta_1$ and $\eta_1$ is a positive constant. Building on Koivumäki et al. (2022), and based on the system's inverse dynamics for the reference SF system in (3) which employs FF compensation $\bar{f}_i$, $x_{id}$ for $i = 2, \ldots, n$ and, control input signal $u(t)$ are defined as follows:

$$\begin{cases} x_{(i+1)d} = \dot{x}_{id} + \bar{f}_i - f_i^* \\ u = \dot{x}_{nd} + \bar{f}_n - f_n^* \end{cases} \tag{13}$$

where FF compensation terms are proposed for each $SS_i$, as:

$$\bar{f}_i = -\frac{1}{2}\gamma_i \bar{e}_i - \epsilon_i \tilde{\theta}_i \frac{\bar{e}_i}{Q_i} - \frac{Q_i}{Q_{i-1}}\bar{e}_{i-1} \tag{14}$$

where $\epsilon_i$, and $\gamma_i$ are positive constants. $Q_i$ is a positive notation, and $\bar{e}_0 = 0$. $\tilde{\theta}_i$ are adaptive laws and can be defined as:

$$\dot{\tilde{\theta}}_i = -\kappa_i \tilde{\theta}_i + \epsilon_i \left(\frac{\bar{e}_i}{Q_i}\right)^2 \tag{15}$$

where $\kappa_i$ is a positive constant. From (12) and (13):

$$\begin{cases} \dot{\bar{e}}_i = \dot{\hat{x}}_i - \dot{x}_{id} = \hat{x}_{i+1} + f_i^* - x_{(i+1)d} + \bar{f}_i - f_i^* \\ \dot{\bar{e}}_n = \dot{\hat{x}}_n - \dot{x}_{nd} = \alpha(u) + f_n^* - u + \bar{f}_n - f_n^* \end{cases} \tag{16}$$

Therefore:

$$\begin{cases} \dot{\bar{e}}_i = \bar{e}_{i+1} + \bar{f}_i \\ \dot{\bar{e}}_n = \Delta_u + \bar{f}_n \end{cases} \tag{17}$$

As observed in (17), $\Delta_u$ is the uncertainty of the reference system when the control input signal exceeds the limitation $u_{min} \le u \le u_{max}$.

### 3.3. Prescribed tracking control performance

In prescribed tracking performance control, singularities are often used to create a constraint that forces the system's state to stay within a desired range of behaviors. This is done by incorporating a barrier function or transformation in the control law, which makes it impossible for the state to approach undesirable values, like going outside a predefined boundary. Built on Zhang and Yang (2019), the expected tracking performance for each $SS_i$ is predetermined by $-o_i(t) < \bar{e}_i < o_i(t)$, where

$$o_i = \left(o_i^{shoot} - o_i^{bound}\right) e^{-o_i^* t} + o_i^{bound} \tag{18}$$

with $o_i^{shoot} > o_i^{bound} > 0$ and $o_i^* > 0$. The overshoot of $\bar{e}_i$ is limited to a value smaller than $o_i^{shoot}$. The final bound and convergence rate of $\bar{e}_i$ are represented by $o_i^{bound}$ and $o_i^*$, respectively. The tracking error-bound $o_i$ should be larger than the bound of the desired reference $\eta_i$, where for each $SS_i$, the domain $-\eta_i \le x_{id} \le \eta_i$ is assumed, while $\eta_i$ is a positive constant. Later in Section 3.4, a conditional function will be described for each $SS_i$ using the properties of $\log\left(\frac{o_i^2}{Q_i}\right)$, whose domain is restricted by the condition that the denominator $Q_i$ must be positive. Let us define

$$Q_i = o_i^2 - \bar{e}_i^2 \tag{19}$$

By selecting $\bar{e}_i(t_0) < o_i(t_0)$, as $\bar{e}_i$ approaches $o_i$, the value inside the logarithm approaches infinity because the denominator approaches zero and the function would be singular and undefined due to a negative argument in the logarithm.

### 3.4. Stability of HAC-BLFs

Motivated by a key concept in the stability connector proposed in Koivumäki et al. (2022) and the virtual power flow proposed in Zhu (2010), for system (12) with the proposed control, new stability connectors are defined as

$$\bar{s}_i = \frac{e_i}{Q_i} e_{i+1} \tag{20}$$

This term is defined to capture the triangular dynamic interactions between adjacent subsystems. Then, this auxiliary function is employed in the convergence analysis of Theorem 3.1 (see Appendix B), demonstrating that the destabilizing interconnections effectively cancel out and that the entire system achieves exponential stability. The error variables $e_i := x_i - \hat{x}_i$ formulated in Section 2 are the differences between the original system signals and the transformed variable. Now, the error variables $\bar{e}_i := \hat{x}_i - x_{id}$ formulated here are the differences between the transformed variable and the reference signals. Based on the definitions of $o_i$, $Q_i$, and $\tilde{\theta}_i$ in Eqs. (18). (19), and (15), respectively, the following quadratic function is defined (Ren, Ge, Tee, & Lee, 2010; Tee, Ge, & Tay, 2009):

$$\bar{V}_{cont} = \sum_{i=1}^{n} \frac{1}{2} \log\left(\frac{o_i^2}{Q_i}\right) + \frac{1}{2}\tilde{\theta}_i^2 \tag{21}$$

Assume that there are any positive parameters $\bar{\rho}_{cont}$, depending on the design parameters of the HAC-BLFs as $\bar{\rho}_{cont} = \min[\min_{i=1}^{n}\{\gamma_i, 2\kappa_i\}]$, and $\bar{\ell}$, depending on the intensity of saturation $\alpha_{max}$ as $\bar{\ell} = \frac{\alpha_{max}^2}{2\upsilon_n Q_n}$. $\upsilon_n$ is any positive constant. Now, the following theorem is stated.

**Proof.** See Appendix B.

**Theorem 3.1.** *Based on (21), employing (14)–(16) for any initial time $t_0$ guarantees that the logarithmic tracking error function $\sum_{i=1}^{n} \log\left(\frac{o_i^2}{Q_i}\right)$ converges exponentially with rate $\bar{\rho}_{cont}$ into a stability region of radius $2\bar{\ell}\bar{\rho}_{cont}^{-1}$ centered at zero. In other words,*

$$\sum_{i=1}^{n} \log\left(\frac{o_i^2}{Q_i}\right) \le 2\bar{V}_{cont}\left(t_0\right) e^{-\left\{\bar{\rho}_{cont}(t-t_0)\right\}} + 2\bar{\ell}\bar{\rho}_{cont}^{-1} \tag{22}$$

### 3.5. Parameter sensitivity analysis

Similar to Section 2.4, the HAC-BLFs provided in (14) for each $SS_i$ require selecting three positive parameters: $\epsilon_i$, $\kappa_i$, and $\gamma_i$. There are three positive parameters forming the stable region: $\bar{V}_{cont}(t_0)$, $\bar{\ell}$, and $\bar{\rho}_{cont}$–which are directly affected by HAC-BLF parameters shown in Table 2, addressing the relevant equations. For further information, see Appendix B. Thus, depending on application demands, and the intensity of the over-saturation control signal, the trade-off between real-time responsiveness and robustness capability based on Table 2 should be considered. Note that, for $i = 1, \ldots, n-1$, $\upsilon_i$ is zero and for $i = n$, $\upsilon_n$ is an arbitrary positive parameter [see (63)].

**Table 2**
Parameter setting of HAC-BLFs, in terms of responsiveness and robustness.

| HAEs' design parameters | Affected parameters | Increase the parameter | Decrease the parameter |
| --- | --- | --- | --- |
| $\epsilon_i$ | None | Insensitive | Insensitive |
| $\min_{i=1}^{n}\{\gamma_i - \upsilon_i,\ 2\kappa_i\}$ | $\bar{\rho}_{cont}$ in (66) | Higher response | More robust |

## 4. Stability of the MRBC framework

The error variables $e_i := x_i - \hat{x}_i$ formulated in Section 2 are the differences between the original system signals and the transformed variable. The error variables $\bar{e}_i := \hat{x}_i - x_{id}$ formulated in Section 3 are the differences between the transformed variable and the reference signals. However, the primary interest is the differences between the original system signals and the transformed variable as $e_i + \bar{e}_i := x_i - x_{id}$. Hence, based on the definitions of $o_i$, $Q_i$, $\tilde{\theta}_i$, and $\tilde{\Psi}_i$ in Eqs. (18). (19), (15), and (5), respectively, the following cumulative quadratic function is defined:

$$V_{all} = V_{ob} + \bar{V}_{cont} = \sum_{i=1}^{n} \frac{1}{2}\left[e_i^2 + \tilde{\Psi}_i^2 + \log\left(\frac{o_i^2}{Q_i}\right) + \tilde{\theta}_i^2\right] \tag{23}$$

Now, the following theorem is stated.

**Theorem 4.1.** *Using the MRBC network, the accumulative error function $\sum_{i=1}^{n} e_i^2 + \log\left(\frac{o_i^2}{Q_i}\right)$ for system (2), facing uncertainties $d_i$, $g_i$, and $\Gamma_i$, over-saturation control signal $\Delta_u$, and PPC $-o_i < \bar{e}_i < o_i$ performance, converges uniformly with an exponentially decay rate $\rho_{all} = \min[\rho_{ob}, \bar{\rho}_{cont}]$, depending on the design parameters of MRBC ($\lambda_i$, $\gamma_i$, $\beta_i$, and $\kappa_i$), into a stability region centered at zero with radius $\ell_{all} = \ell_{ob} + \bar{\ell}$ depending on the intensity of the uncertainties.*

**Proof.** See Appendix C.

## 5. Experimental validity

The MRBC framework – including both HAEs and HAC-BLFs – was implemented on a real-world system testbed with SF form: an electromechanical linear actuator (EMLA) system which converts the rotary motion of an electric motor into high-force linear displacement via a mechanical transmission such as a ball screw. This actuator prototype will be fixed to the joint base of a heavy-duty robotic manipulator, and its moving nut (or carriage) is rigidly attached to the link of the manipulator. The setup under study, along with the MRBC framework, is shown in Fig. 3, while a link to the open-source demonstration video is available in the caption of the figure. The studied EMLA setup (see Fig. 4) features an 8-pole, three-phase Nidec PMSM (permanent-magnet synchronous motor) operating at 380/480 V and rated for 11.6 kW. A Unidrive M700 controller manages the motor, providing a subsystem-level interface for inverter switching and torque control. The motor's output shaft drives a reduction gearbox, which in turn powers a ball screw assembly, transforming the motor's rotary motion into accurate linear movement. In addition, a hydraulic cylinder is coupled to the EMLA's output to emulate varying external loads.

Functioning as an active force emulator, this hydraulic unit produces specified force trajectories through its dedicated HMI. Electro-hydraulic control valves adjust the force in real time, reproducing the time-dependent resistance encountered at a heavy-duty manipulator's lift joint. All components – the motor controller, inverter, load sensors, and hydraulic system – are synchronized over a real-time EtherCAT network, guaranteeing high-speed data exchange and precise actuation feedback.

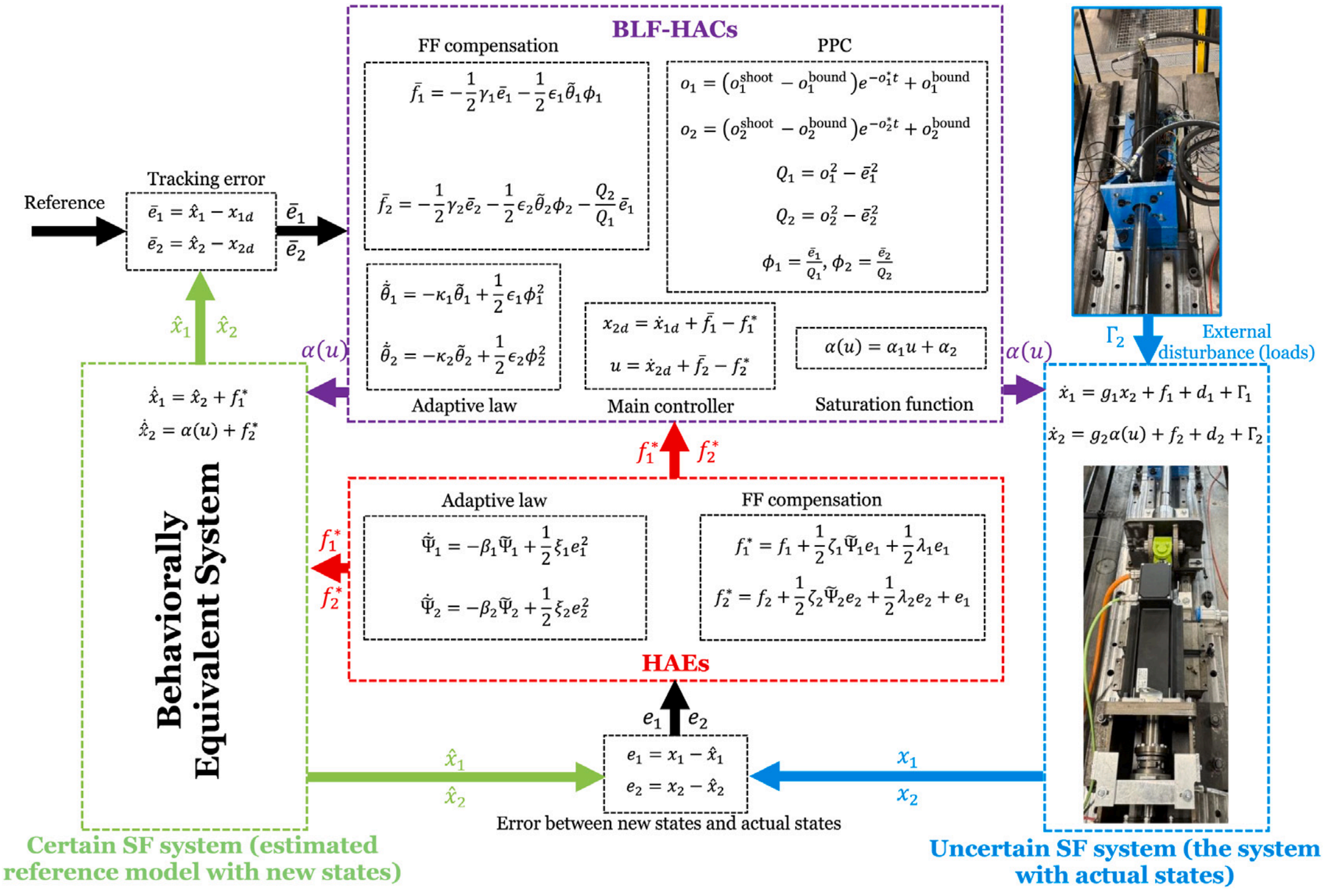

**Fig. 3.** Schematic of the MRBC framework applied to the studied EMLA. The relevant video is available at: https://youtu.be/udFBJ0T9d8A.

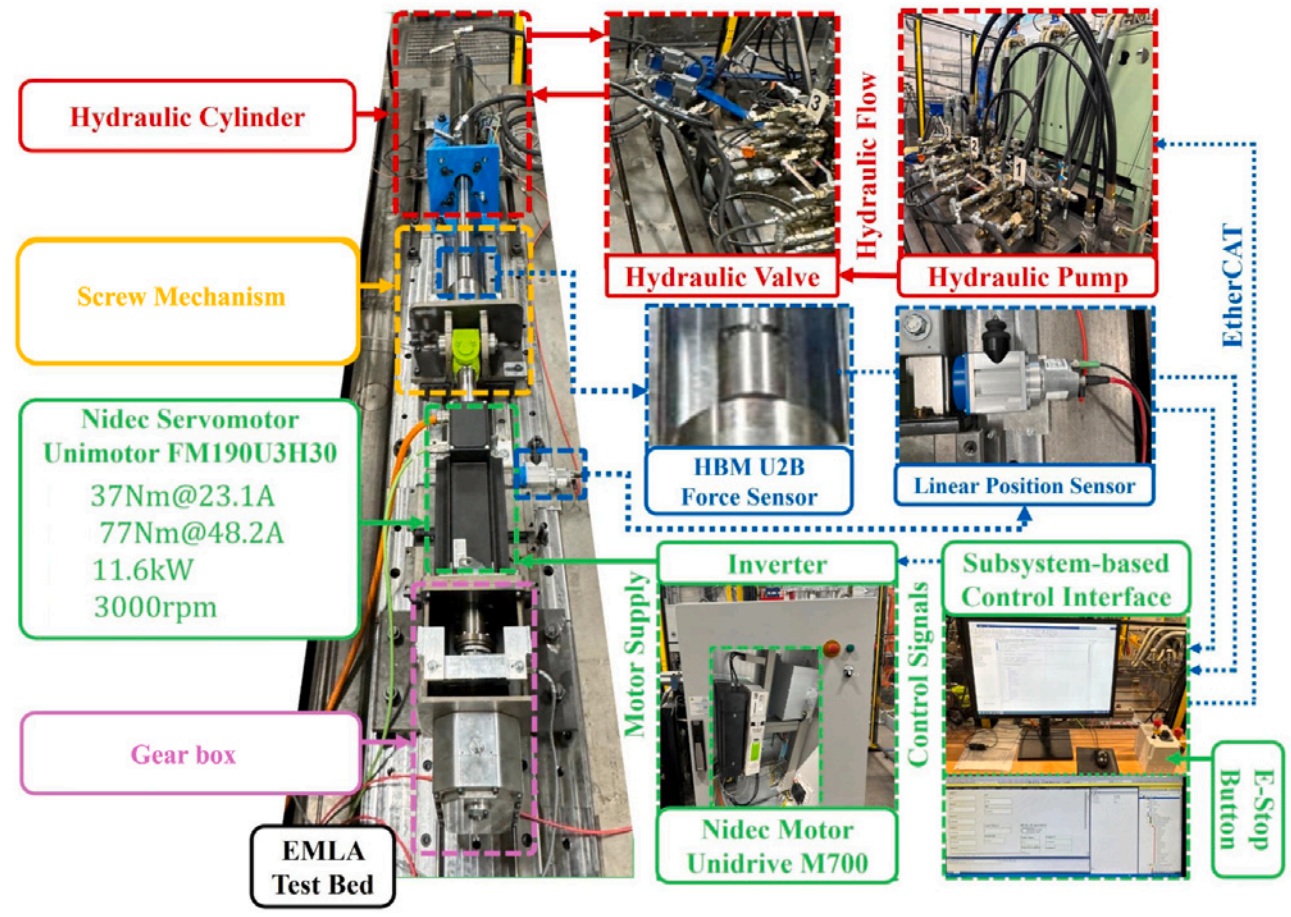

**Fig. 4.** Experiment setup (Bahari, Paz, Shahna, Mustalahti, & Mattila, 2025).

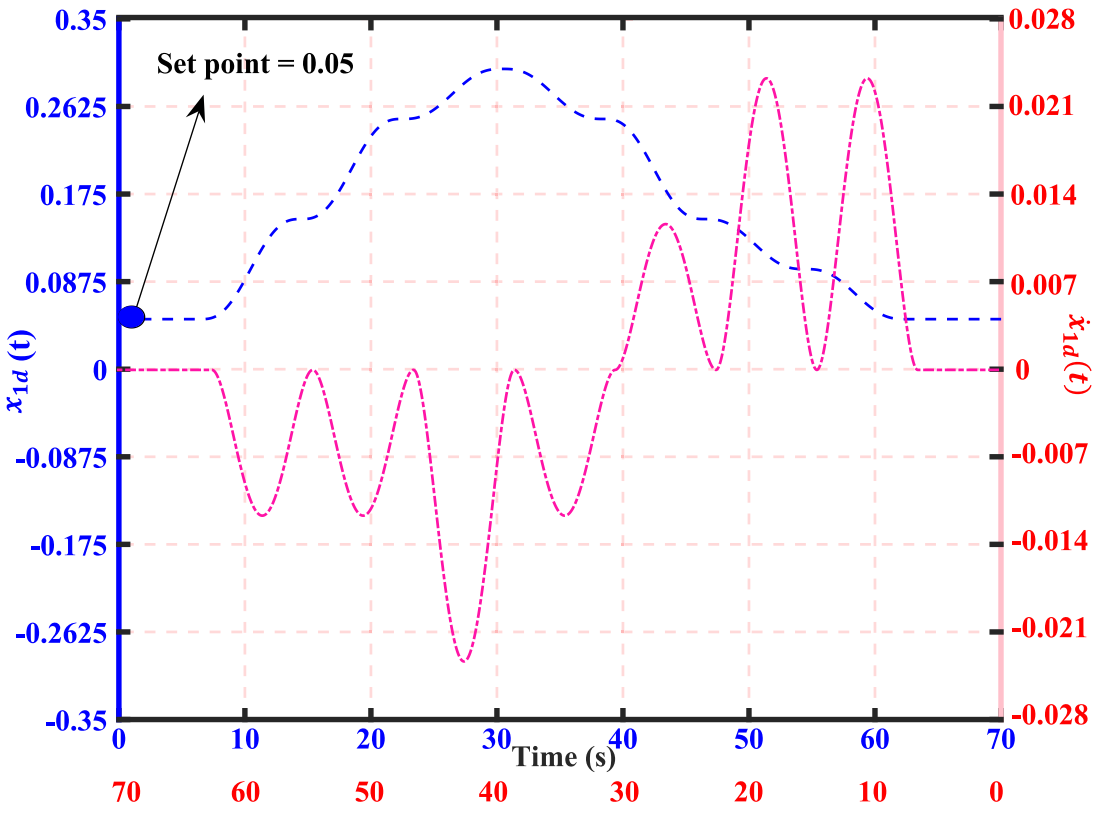

**Fig. 5.** Reference Trajectory for experiments 1 to 4: $x_{1d}$ based on quantic polynomials (Jazar, 2010) with set point 0.05 and its derivative $\dot{x}_{1d}$.

The uncertain SF system represents the actual EMLA, incorporating unknown disturbances and nonlinearities introduced by the hydraulic linear actuator. On the left, the certain reference model – with equivalent system states to EMLA states – is constructed by estimating new model terms $f_1^*$ and $f_2^*$ through the HAEs, resulting in a behaviorally equivalent system that replicates the input–output dynamics of the real system. These estimated functional modeling terms are then provided to both the BLF-HAC block and the reference tracking error. The proposed controller, which includes FF compensation, adaptive laws, and PPCs, is designed to generate sufficient motor torque to ensure system stability and achieve the desired tracking performance under predefined constraints. The motion dynamics of the studied EMLA mechanism can be represented as an uncertain SF system, based on Shahna, Mustalahti and Mattila (2024), as follows:

$$\dot{v}_L = I_{eq}^{-1}(\tau_m - B_{eq}v_L - K_{eq}x_L - f_{eq}F_L) \tag{24}$$

where $\tau_m$, $I_{eq}, B_{eq}, K_{eq}$, $f_{eq}$, and $F_L$ are the motor torque (measured in N m), equivalent inertia (kg m$^2$), damping ($\frac{\text{N s}}{\text{m}}$), spring effect ($\frac{\text{N}}{\mu\text{ m}}$), load coefficient of the actuator system, and load disturbance (N), respectively. By defining the position state $x_L$ (m) and velocity state $\dot{x}_L = v_L$ ($\frac{\text{m}}{\text{s}}$) as $x_1$ and $x_2$, respectively, there is the following representation, based on (2):

$$\begin{cases} \dot{x}_1 = g_1\left(\boldsymbol{x}_1, t\right) x_2 + f_1\left(\boldsymbol{x}_1, t\right) + d_1\left(\boldsymbol{x}_1, t\right) + \Gamma_1 \\ \dot{x}_2 = g_2\left(\boldsymbol{x}_2, t\right) u + f_2\left(\boldsymbol{x}_2, t\right) + d_2\left(\boldsymbol{x}_2, t\right) + \Gamma_2 \end{cases} \tag{25}$$

where the modeling term for $SS_1$ is $f_1 = 0$. However, $g_1$, $d_1$, and $\Gamma_1$ are related to sensory noise and the derivative effects of sensor data, causing an undesired and nonlinear relationship between the derivative of position $\dot{x}_1$ and velocity $x_2$, for which specific information was lacking. Similarly, for $SS_2$, there are $u = \tau_m$, $g_2 = I_{eq}^{-1}$, $f_2 = -I_{eq}^{-1}K_{eq}x_L$, and $d_2 = -I_{eq}^{-1}B_{eq}v_L$. Note that $g_{1,2}$, $d_{1,2}$, and $\Gamma_{1,2}$ are unknown for the MRBC framework. As mentioned, an additional actuator was also employed at the load side of the experimental system to generate functional loading force $F_L$, which introduced an unknown load disturbance, $\Gamma_2 = -I_{eq}^{-1} f_{eq} F_L$, on the studied EMLA, varying its motor torque capacity by 0–95%. The control signals and communications between the setup components were managed via an EtherCAT network operated at a sampling rate of 1000 Hz.

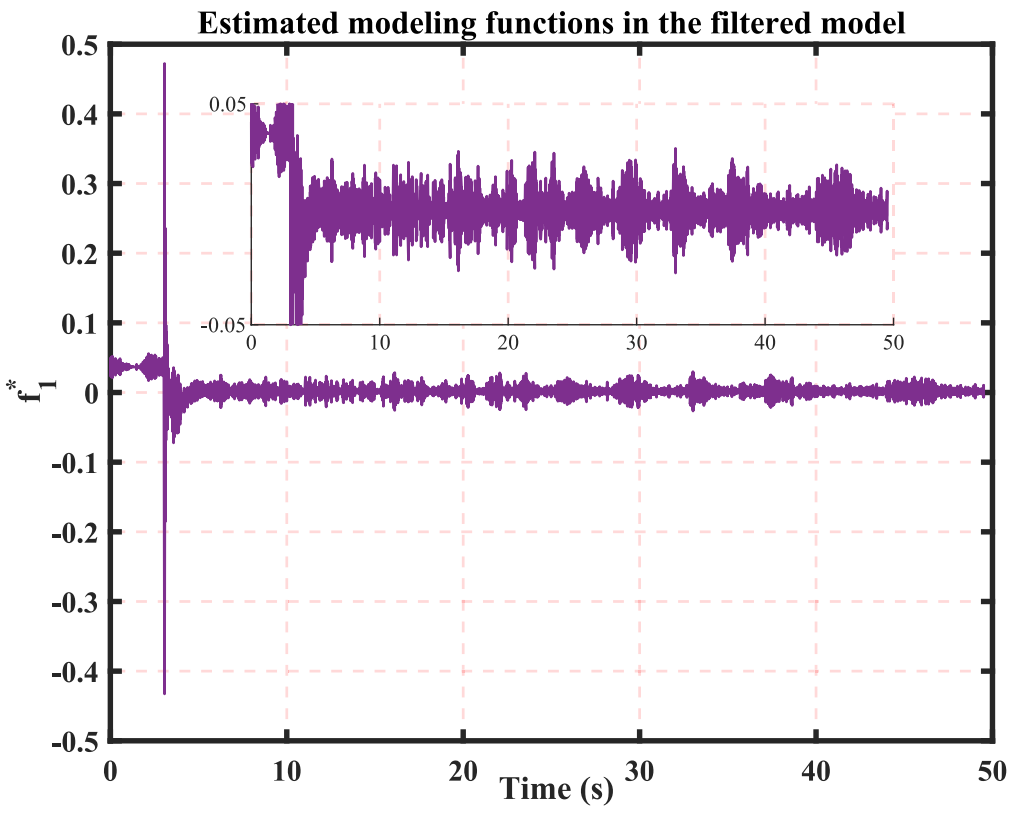

**Fig. 6.** Experiment 1: the values of $f_1^*$ (measured in N m).

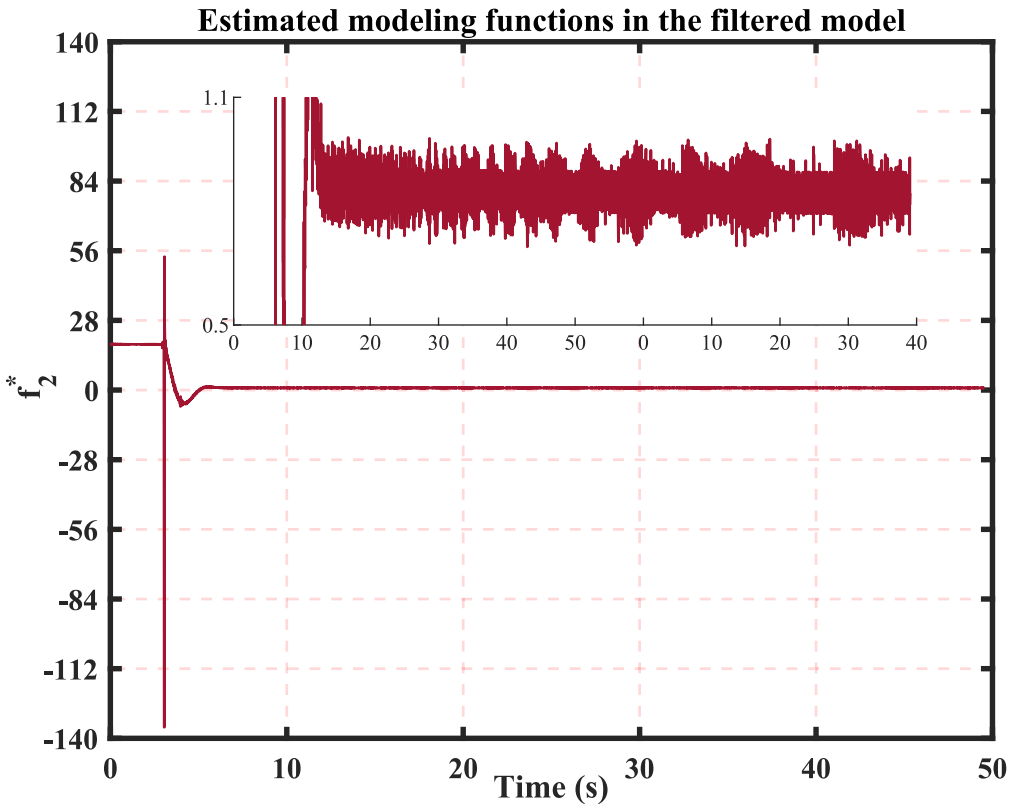

**Fig. 7.** Experiment 1: the values of $f_2^*$ (measured in N m).

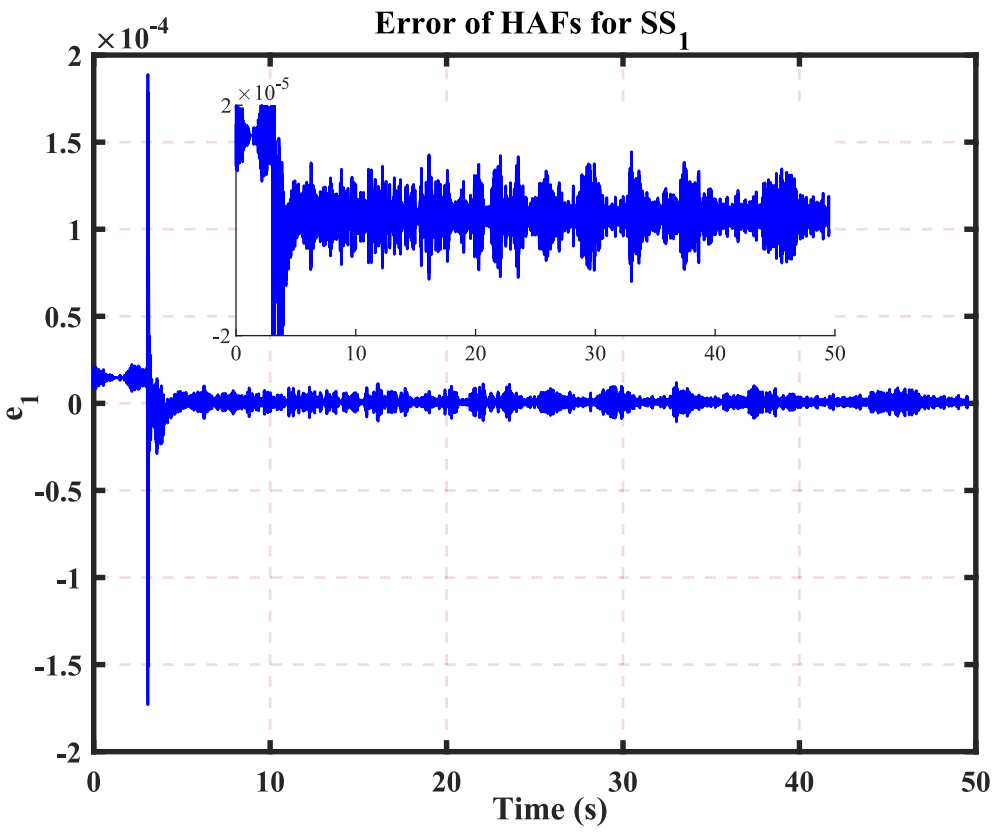

**Fig. 8.** Experiment 1: errors between the new state of the certain model (3) estimated by employing HAEs and the state of the initial system (2) for $SS_1$.

Two set points at initial time $t_0$ were considered as $x_{1d}(t_0) = 0.05$ m and $\dot{x}_{1d}(t_0) = 0$ m/s for the system sustainability under varying load disturbances. Then, reference trajectories are defined based on quintic polynomials from Jazar (2010) for the system tracking under varying load disturbances. From (13), the reference setpoint/trajectory of $SS_1$ as $x_{1d}$ and for $SS_2$ as $x_{2d} = \dot{x}_{1d} + \bar{f}_1 - f_1^*$ are obtained.

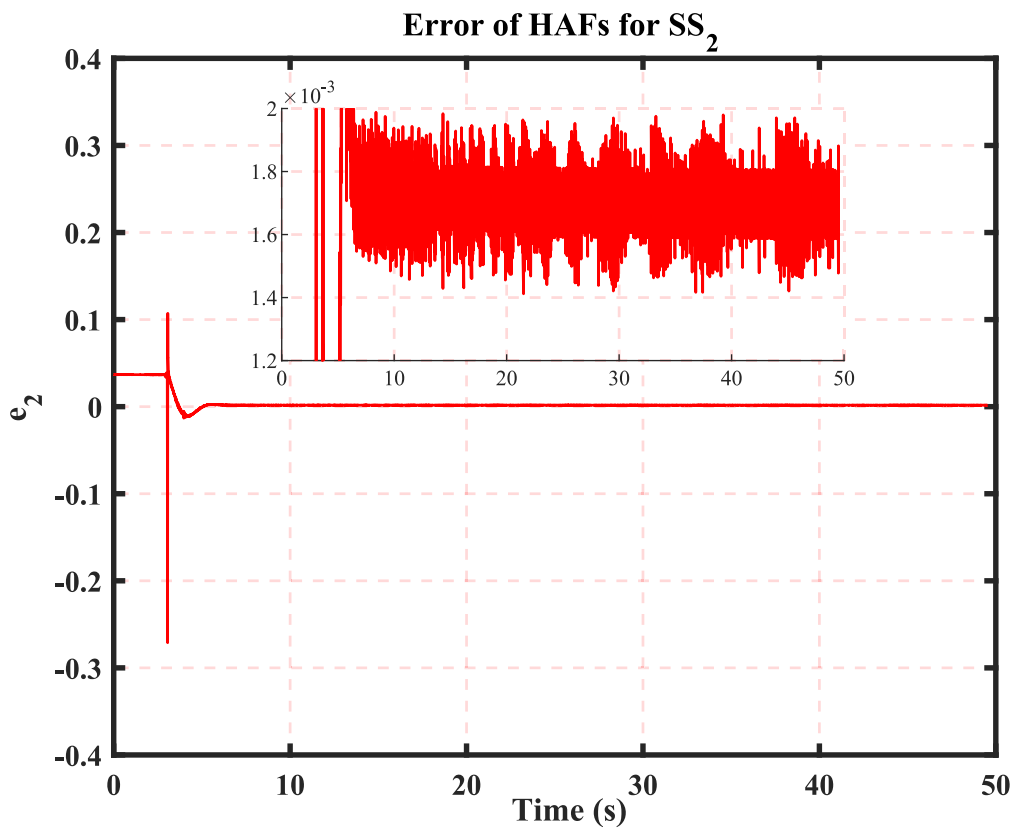

**Fig. 9.** Experiment 1: errors between the new state of the certain model (3) estimated by employing HAEs and the state of the initial system (2) for $SS_2$.

Control parameters of the MRBC can be tuned systematically or via trial-and-error, considering sensitivity analysis and the trade-off between robustness and responsiveness based on application-specific disturbance intensity, as discussed in Sections 2.3 and 3.5. In this work, the control parameters were tuned by employing a heuristic optimization method – the Jaya algorithm (Shahna & Mattila, 2024) – to determine the suboptimal control gains. The Jaya algorithm was initiated by generating an initial population the vector of control gains, denoted as $(c)$ through random sampling. For each candidate in this population, a cost function was evaluated based on the standard deviation of the position and velocity tracking errors, which served as the objective to be minimized. The best-performing candidate, denoted as $c_{\text{best}}$, was identified by the lowest cost value, while the worst-performing candidate, $c_{\text{worst}}$, corresponded to the highest cost value. These values were then used to iteratively update the population and generate a new candidate solution, $c_{\text{new}}$, according to the following update rule:

$$c_{new} = c + r_1 \left( c_{best} - c \right) - r_2 \left( c_{worst} - c \right) \tag{26}$$

By doing this, the parameters of HAEs were set as $\lambda_{1,2} = 500$, $\beta_{1,2} = 0.8$, and $\xi_{1,2} = 0.08$, while the control parameters of HAC-BLFs were set as $\gamma_{1,2} = 40$, $\epsilon_{1,2} = 0.8$, and $\kappa_{1,2} = 1$. As shown in Tables 1 and 2, the values of $\xi_i$ and $\epsilon_i$ do not significantly affect the outcome. However, depending on the intensity of the disturbances, a trade-off adjustment occurs between $\lambda_i$ and $\beta_i$, as well as between $\kappa_i$ and $\gamma_i$.

To validate the proposed MRBC framework, five experiments were conducted on the studied EMLA system, as follows: (1) The EMLA stabilizes at the set point $x_{1d} = 0.05$ m within the predefined control performance specifications under low-disturbance conditions (with motor capacity variations of up to 12.5%). (2) It stabilizes at the set point $x_{1d} = 0.05$ m within the predefined control performance specifications under high-disturbance conditions (with motor capacity variations of up to 92%). (3) It tracks the reference trajectory $x_{1d}$ shown in Fig. 5 within the predefined control performance specifications under low-disturbance conditions (with motor capacity variations of up to 13%). (4) It tracks the reference trajectory $x_{1d}$ shown in Fig. 5 within the predefined control performance specifications under high-disturbance conditions (with motor capacity variations of up to 95%). (5) The proposed MRBC framework is compared against the adaptive asymptotic tracking control (AATC) proposed in Zhao, Song, Chen, and Chen (2021) and the adaptive fault-tolerant control (AFTC), for the studied EMLA system tracking a repetitive trapezoidal reference profile, despite the presence of parameter uncertainties in the load. In this profile, the EMLA is commanded to hold its initial position at 5 cm for 5 s.

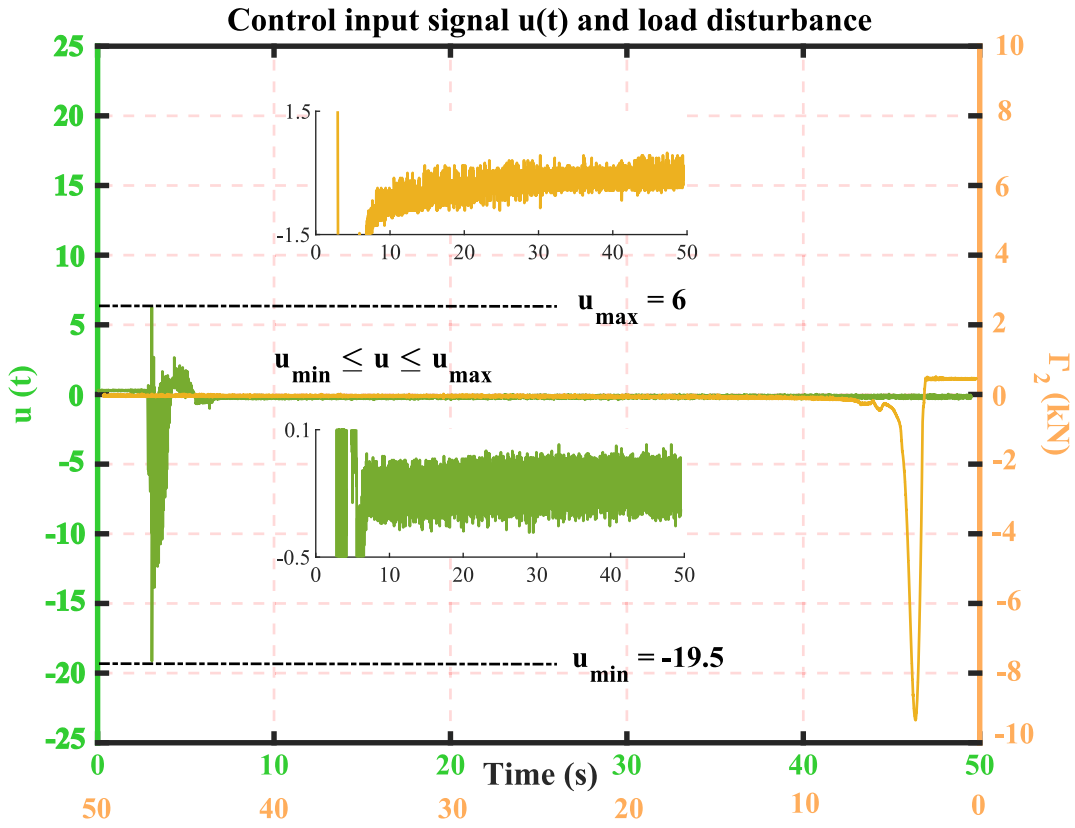

**Fig. 10.** Experiment 1: the control signal constrained by the function $\alpha(.)$ and external force $F_L$ imposed on the system. (For interpretation of the references to color in this figure legend, the reader is referred to the web version of this article.)

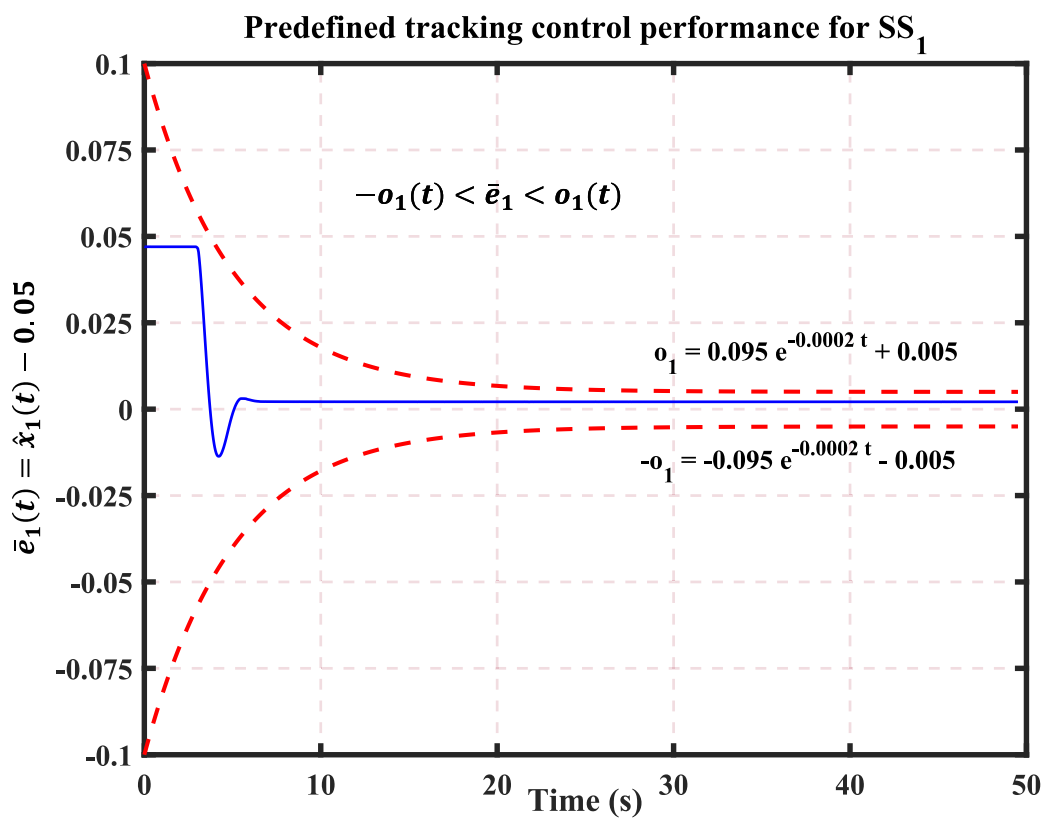

**Fig. 11.** Experiment 1: Predefined set point-reaching performance of $\mathrm{SS}_1$.

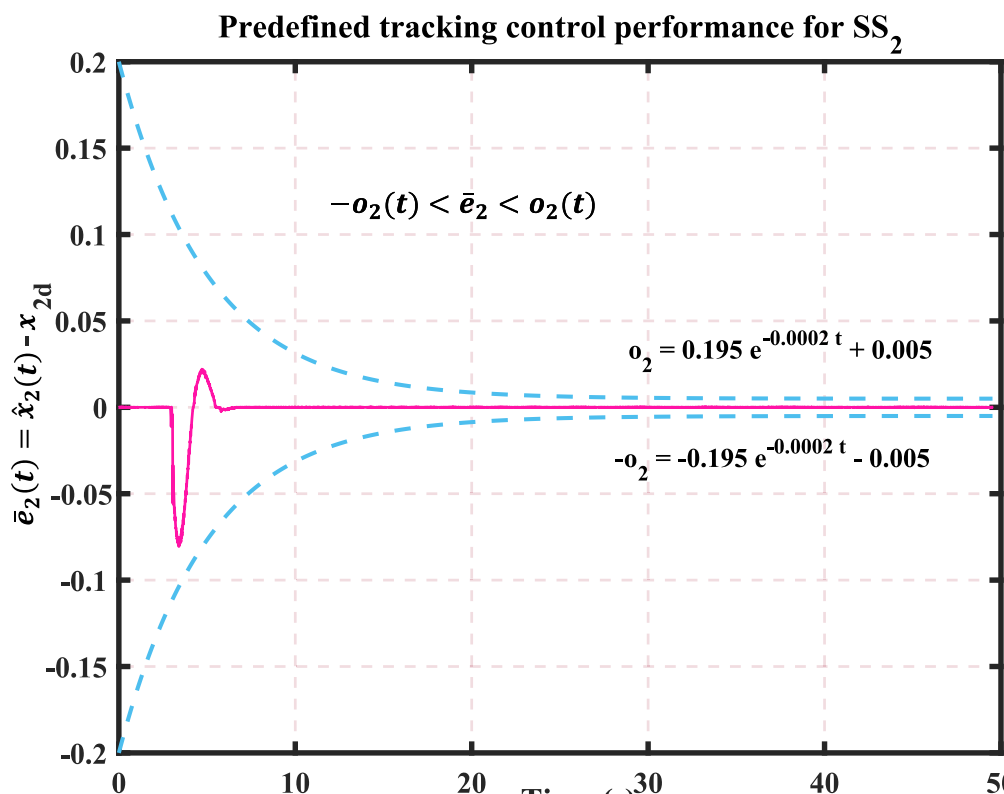

**Fig. 12.** Experiment 1: predefined control performance of $\mathrm{SS}_2$.

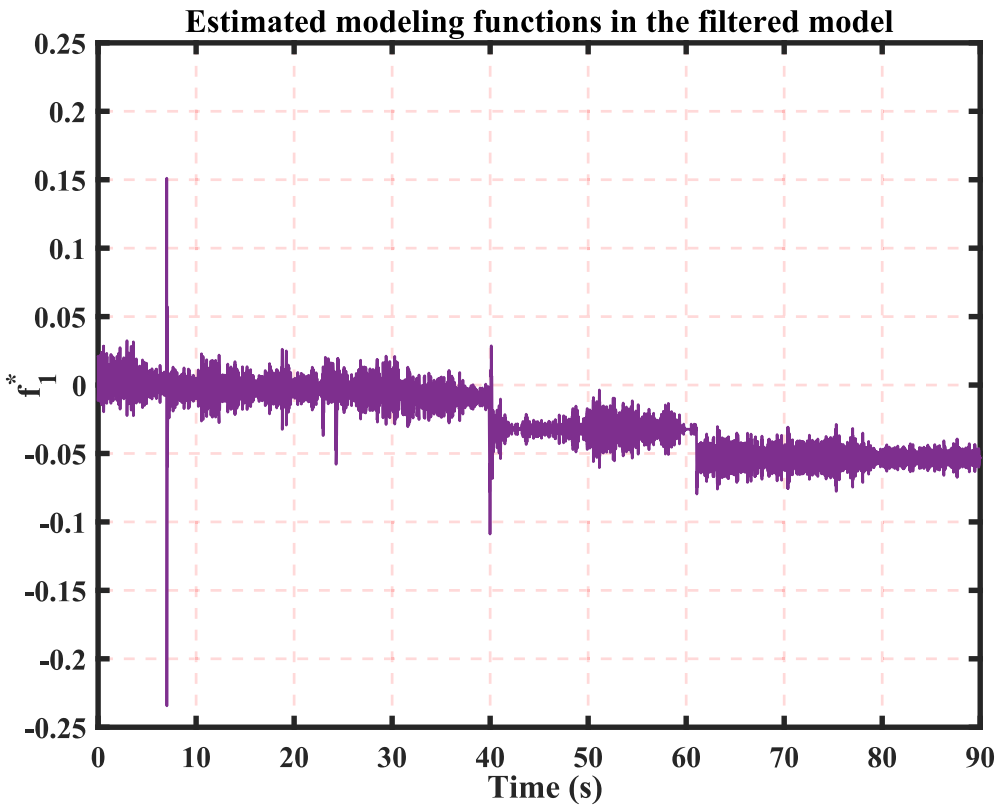

**Fig. 13.** Experiment 2: the values of $f_1^*$ (measured in N m).

It then moves forward with a gradually varying velocity—accelerating up to 5 cm/s and then decelerating back to zero. Upon reaching the target position of 25 cm, it holds that position for 10 s. The system then returns to the initial position following the same velocity profile in reverse. This cycle repeats periodically.

### 5.1. Experiment 1: Set point stabilization under 0–12.5% of the EMLA's capacity

Regardless of the system's initial condition, the objective was to drive the system to reach and maintain the setpoint of $x_{1d} = 0.05$, despite uncertainties in system and control directions ($d_{1,2}$ and $g_{1,2}$), as well as external disturbances ($\Gamma_{1,2}$). Based on Section 2, and employing HAEs, the objective was to estimate $f^*_{1,2}$ and transform the uncertain system (25) into the certain model system (3), as follows:

$$\begin{cases} \dot{\hat{x}}_1(t) = \hat{x}_2(t) + f_1^*\left(x_1, t\right) \\ \dot{\hat{x}}_2(t) = u(t) + f_2^*\left(\boldsymbol{x}_2, t\right) \end{cases} \tag{27}$$

Figs. 6 and 7 illustrate the estimated values of $f_1^*$ in purple and $f_2^*$ in dark red on two different scales (both measured in N m) in the reference system. The errors between the new states of the reference model in (27) and the actual states of (25), are illustrated in Figs. 8 and 9 as $e_1 = x_1 - \hat{x}_1$ for $\mathrm{SS}_1$ in blue and $e_2 = x_2 - \hat{x}_2$ for $\mathrm{SS}_2$ in red.

The newly estimated states, $\hat{x}_1$ and $\hat{x}_2$, were obtained separately from the HAE's input HAC-BLFs. To stabilize the states of the reference system $\hat{x}_1$ at the set point ($x_{1d} = 0.05$), from (12), the reference trajectory for $\mathrm{SS}_2$ based on the reference set point of $\mathrm{SS}_1$ ($\dot{x}_{1d} = 0$) as $x_{2d} = 0 + \bar{f}_1 - f_1^* = \bar{f}_1 - f_1^*$ was obtained. Similarly the control input signal was obtained as $u = \dot{x}_{2d} + \bar{f}_2 - f_2^*$. Then, following (11), the control signal was constrained as $-19.5 \leq u \leq 6$. The green trajectory in Fig. 10 illustrates the generated control signal, which adheres to the defined constraint. The orange trajectory in this figure also shows the external force $F_L$ applied to the EMLA (extracting 0–12.5% of its capacity) to generate the low-disturbance term $\Gamma_2 = -I_{eq}^{-1} f_{eq} F_L$ in this experiment. Based on (18), the expected control performance for $\mathrm{SS}_1$ was defined as $o_1^{shoot} = 0.1$, $o_1^{bound} = 0.005$, and $o_1^* = 0.0002$. Hence, Fig. 11 illustrates the control error for $\mathrm{SS}_1$ by employing HAC-BLFs, adhering to the PPC within $-o_1(t) < \bar{e}_1 < o_1(t)$. Similarly, the expected control performance for $\mathrm{SS}_2$ was defined as $o_2^{shoot} = 0.2$, $o_2^{bound} = 0.005$, and $o_2^* = 0.0002$. Hence, Fig. 12 illustrates the control error for $\mathrm{SS}_2$ by employing HAC-BLFs, adhering to the PPC within $-o_2(t) < \bar{e}_2 < o_2(t)$.

### 5.2. Experiment 2: Set point stabilization under 0–92% of the EMLA's capacity

In this experiment, similar to the previous one, regardless of the system's initial point, the objective was to have the system reach and maintain the set point of $x_{1d} = 0.05$ under the high-disturbance condition. Again, based on Section 2, and employing HAEs, the objective was to estimate $f^*_{1,2}$ and transform the uncertain system (2) into the certain model system (3). As observed in Figs. 13 and 14, gaining disturbance $\Gamma_2$ in $\mathrm{SS}_2$ did not have a visible effect on $f_1^*$ (purple trajectory) relative to $\mathrm{SS}_1$, while the estimation of $f_2^*$ (dark red trajectory) was impacted prominently. Figs. 15 and 16 show the errors between the new states of the reference model in (3), and the rough states of the initial system (2)($e_1 = x_1 - \hat{x}_1$ for $\mathrm{SS}_1$ in blue and $e_2 = x_2 - \hat{x}_2$ for $\mathrm{SS}_2$ in red). Compared to experiment 1, the errors between the new states of a certain model in (3) and states of the initial form of system (2) under high-disturbance conditions had a better transient but a reduction in steady-state performance. The new estimated states $\hat{x}_1$ and $\hat{x}_2$ of the system were obtained separately from HAEs and input HAC-BLFs. To stabilize the states of the reference system $\hat{x}_1$ at the set point ($x_{1d} = 0.05$), like experiment 1, $x_{2d} = 0 + \bar{f}_1 - f_1^* = \bar{f}_1 - f_1^*$ was obtained, and similarly, the control input signal was obtained as $u = \dot{x}_{2d} + \bar{f}_2 - f_2^*$. Then, the control signal was constrained as $-2.5 \leq u \leq 38$, following (11). As the external disturbance was a loading force about 10 times more intense compared to that in experiment 1, the value of the upper bound of control input $u_{max}$ was considered larger, while the lower bound $u_{min}$ was considered smaller. The green trajectory in Fig. 17 illustrates the generated control signal, which adheres to the defined constraint while its amplitude increase to control the system under high-disturbance conditions. The orange trajectory in this figure also shows the external force $F_L$ applied to the EMLA (extracting 0–92% of its capacity) to generate the external disturbance $\Gamma_2 = -I_{eq}^{-1} f_{eq} F_L$. In this scenario, based on (18), the expected control performance for $SS_1$ was defined as $o_1^{shoot} = 0.005$, $o_1^{bound} = 0.002$, and $o_1^* = 0.0001$. Fig. 18 illustrates the control error incurred for $SS_1$ by employing HAC-BLFs, adhering to the PPC within $-o_1(t) < \bar{e}_1 < o_1(t)$. Similarly, the expected control performance for $SS_2$ was defined as $o_2^{shoot} = 0.150$, $o_2^{bound} = 0.015$, and $o_2^* = 0.0004$. Fig. 19 illustrates the control error incurred for $SS_2$ by employing HAC-BLFs, adhering to the PPC within $-o_2(t) < \bar{e}_2 < o_2(t)$.

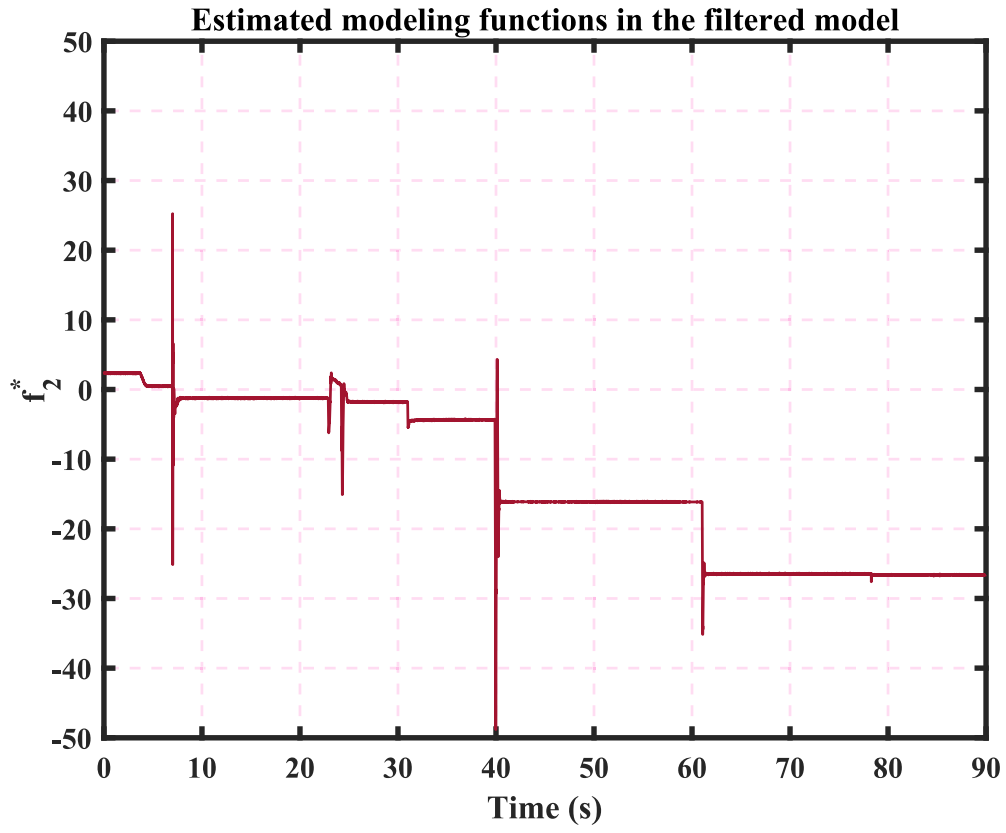

**Fig. 14.** Experiment 2: the values of $f_2^*$ (measured in N m).

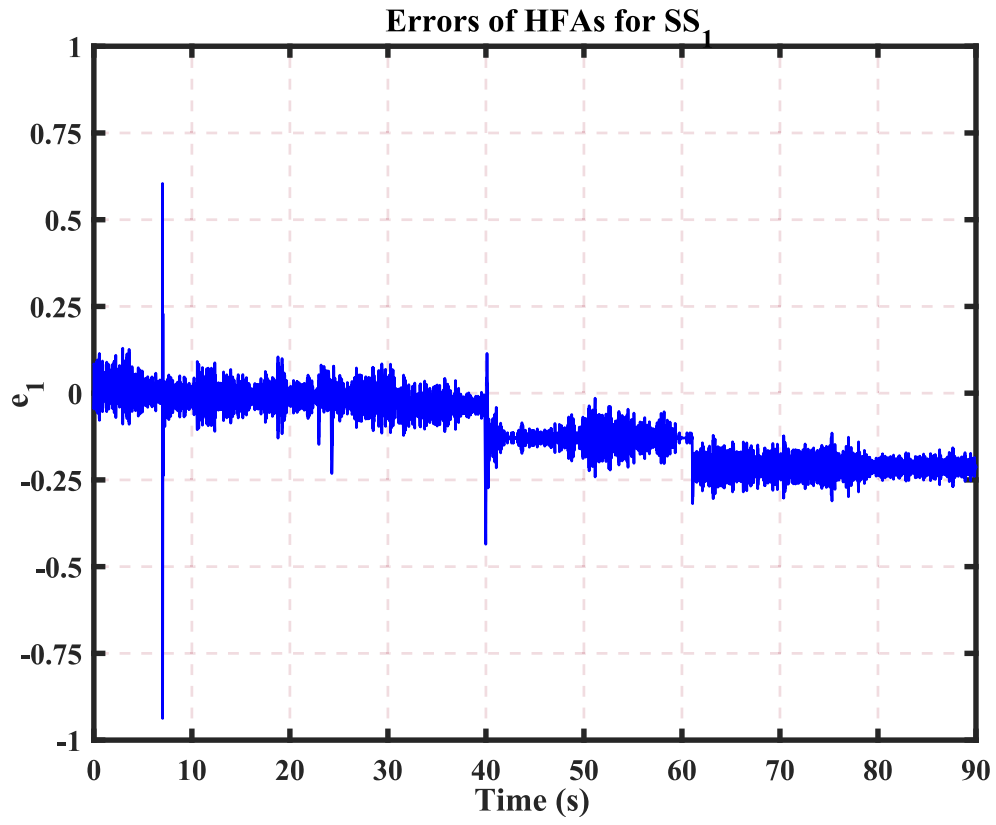

**Fig. 15.** Experiment 2: errors between the new state of the certain model in (3) estimated by employing HAEs and the state of the initial system (2) for $SS_1$.

### 5.3. Experiment 3: Trajectory tracking under 0–13% of the EMLA's capacity

In this experiment, regardless of the system's initial point, the objective was to have it track the reference trajectory $x_{1d}$ shown in Fig. 5 under the low-disturbance conditions similar to the disturbance imposed on the system in experiment 1. Again, based on Section 2, and employing HAEs, the objective was to estimate $f_{1,2}^*$ and transform the uncertain system (2) into the certain model system (3). As observed in Fig. 20, there is no prominent change in $f_1^*$ (purple trajectory) relative to $SS_1$. In contrast, in Fig. 21 the estimation of $f_2^*$ (dark red trajectory) was impacted prominently compared to experiment 1 because of the uncertain term $d_2 = -I_{eq}^{-1} B_{eq} x_2$, generating more intense uncertainties by varying $x_2$, although $f_2^*$ was lower than that in experiment 2 facing high disturbance. Figs. 22 and 23 show the errors between the new states of reference model (3), and the rough states of the initial system (2) ($e_1 = x_1 - \hat{x}_1$ for $SS_1$ in blue and $e_2 = x_2 - \hat{x}_2$ for $SS_2$ in red).

Compared to experiments 1 and 2, the error of state estimation in $SS_1$ was roughly similar, while that in $SS_2$ was worse than experiment 1 and better than experiment 2. The new estimated states $\hat{x}_1$ and $\hat{x}_2$ of the system were obtained from HAEs and input HAC-BLFs, separately. The objective was for $\hat{x}_1$ to track the reference trajectory $x_{1d}$ shown in Fig. 3. Based on (13), the reference trajectory for $SS_2$ was obtained as $x_{2d} = \dot{x}_{1d} + \bar{f}_1 - f_1^*$, and similarly the control input signal was obtained as $u = \dot{x}_{2d} + \bar{f}_2 - f_2^*$. Then, the control signal was constrained, as $-13 \leq u \leq 6$, following (11).

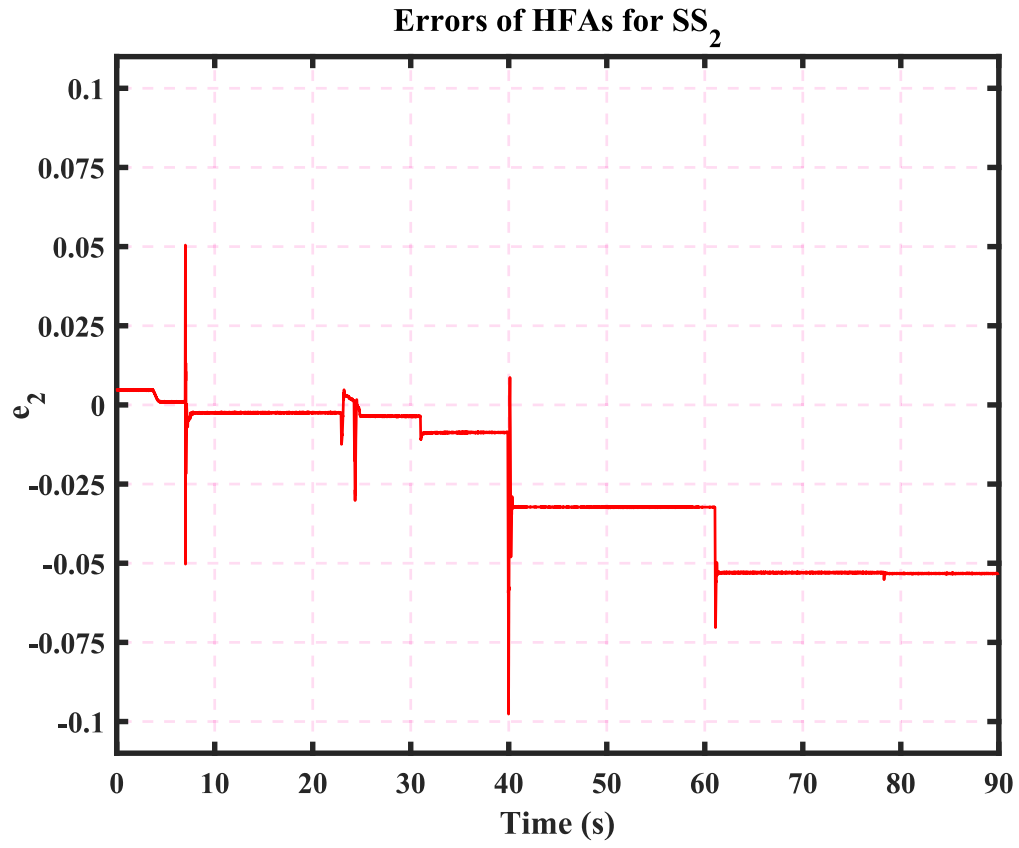

**Fig. 16.** Experiment 2: the errors between the new state of the certain model in (3) estimated by employing HAEs and the state of the initial system in (2) $SS_2$.

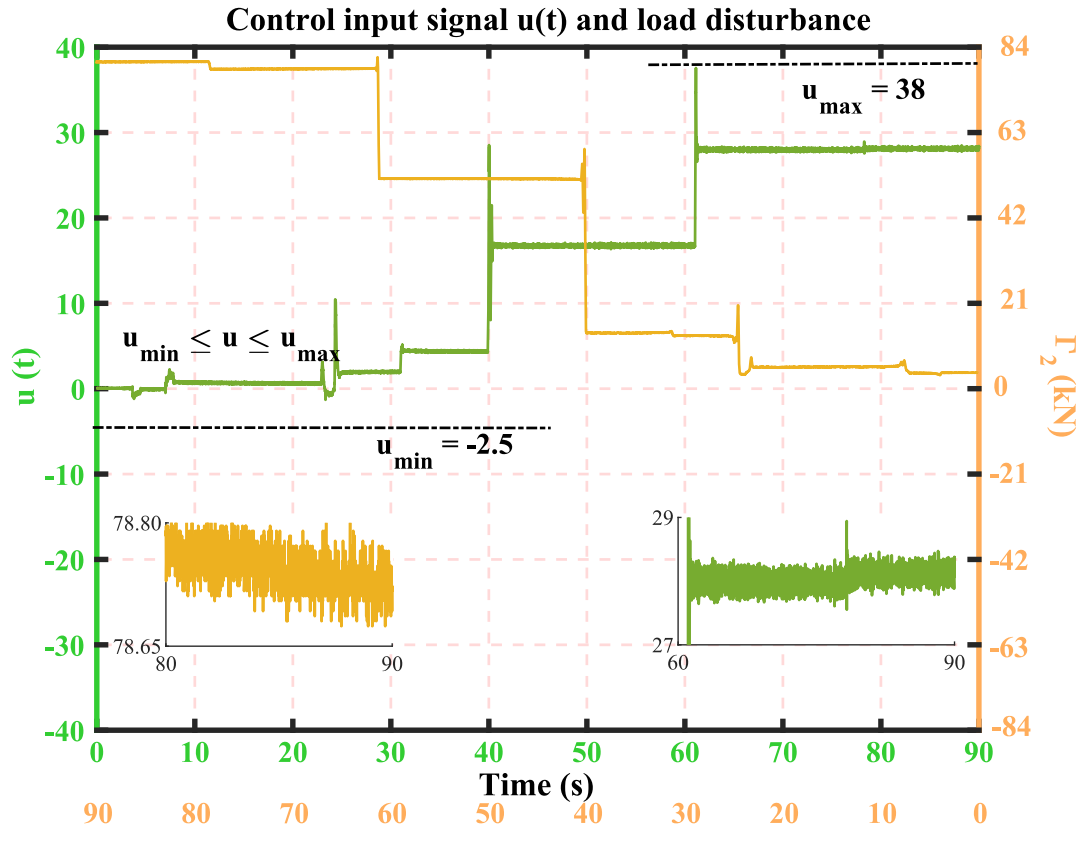

**Fig. 17.** Experiment 2: the control signal constrained by the function $\alpha(.)$ and the external force $F_L$ imposed on the system. (For interpretation of the references to color in this figure legend, the reader is referred to the web version of this article.)

The green trajectory in Fig. 24 illustrates the generated control signal, which adheres to the defined constraint. The orange trajectory in this figure also shows the external force $F_L$ applied to the EMLA (extracting 0–13% of its capacity) to generate the external disturbance $\Gamma_2 = -I_{eq}^{-1} f_{eq} F_L$ in this experiment. Based on (18), the expected control performance for $SS_1$ was defined as $o_1^{shoot} = 0.2$, $o_1^{bound} = 0.015$, and $o_1^* = 0.00018$. Fig. 25 illustrates the control tracking error for $SS_1$ by employing HAC-BLFs, adhering to the PPC within $-o_1(t) < \bar{e}_1 < o_1(t)$. Similarly, the expected control performance for $SS_2$ was defined as $o_2^{shoot} = 0.2$, $o_2^{bound} = 0.02$, and $o_2^* = 0.0002$. Fig. 26 illustrates the control error for $SS_2$ by employing HAC-BLFs, adhering to the PPC within $-o_2(t) < \bar{e}_2 < o_2(t)$.

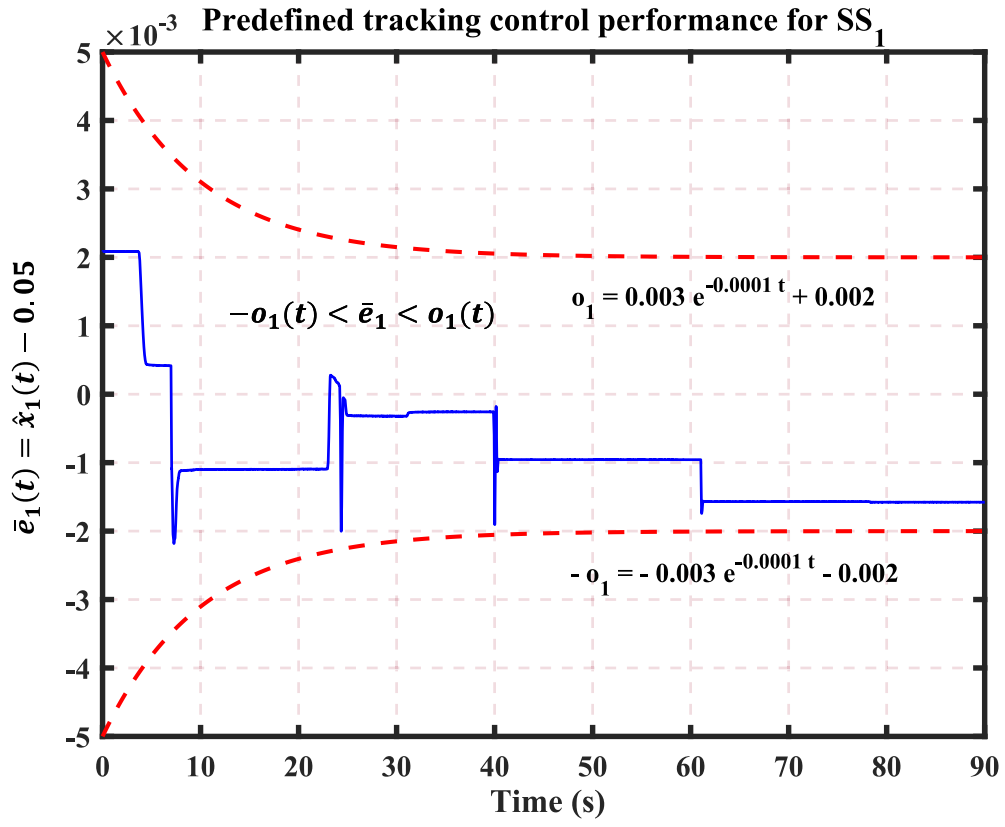

**Fig. 18.** Experiment 2: predefined set point-reaching performance of $SS_1$.

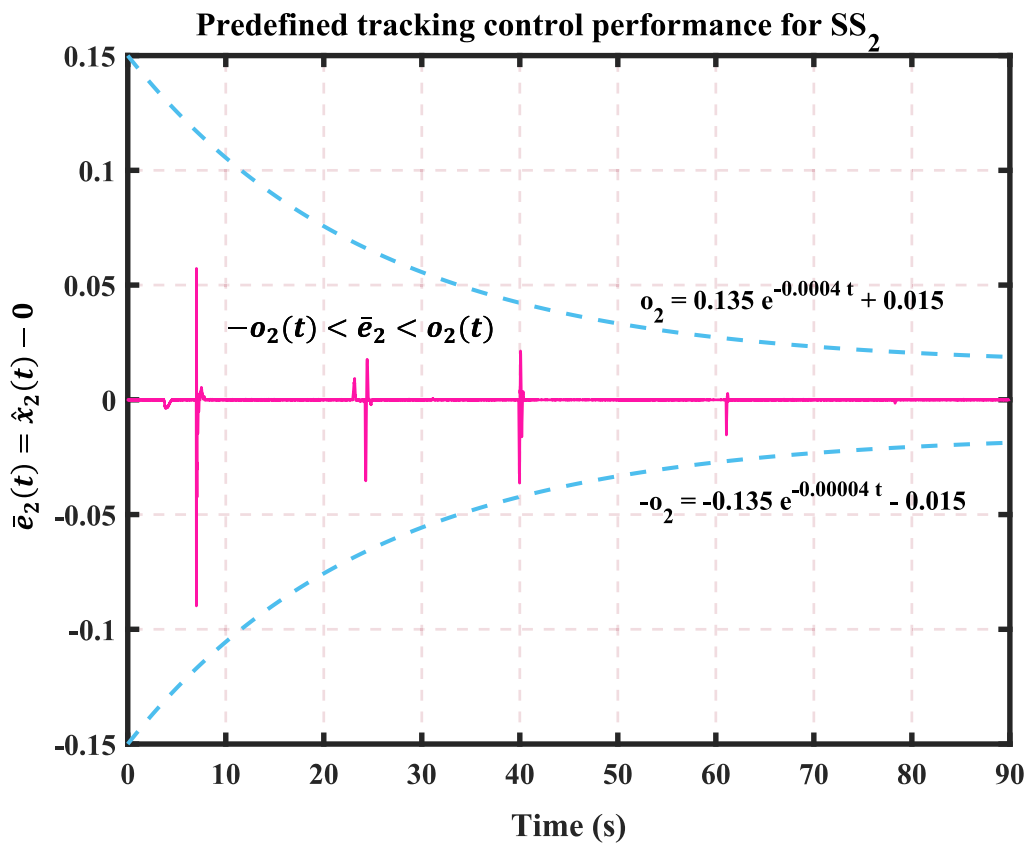

**Fig. 19.** Experiment 2: predefined control performance of $SS_2$.

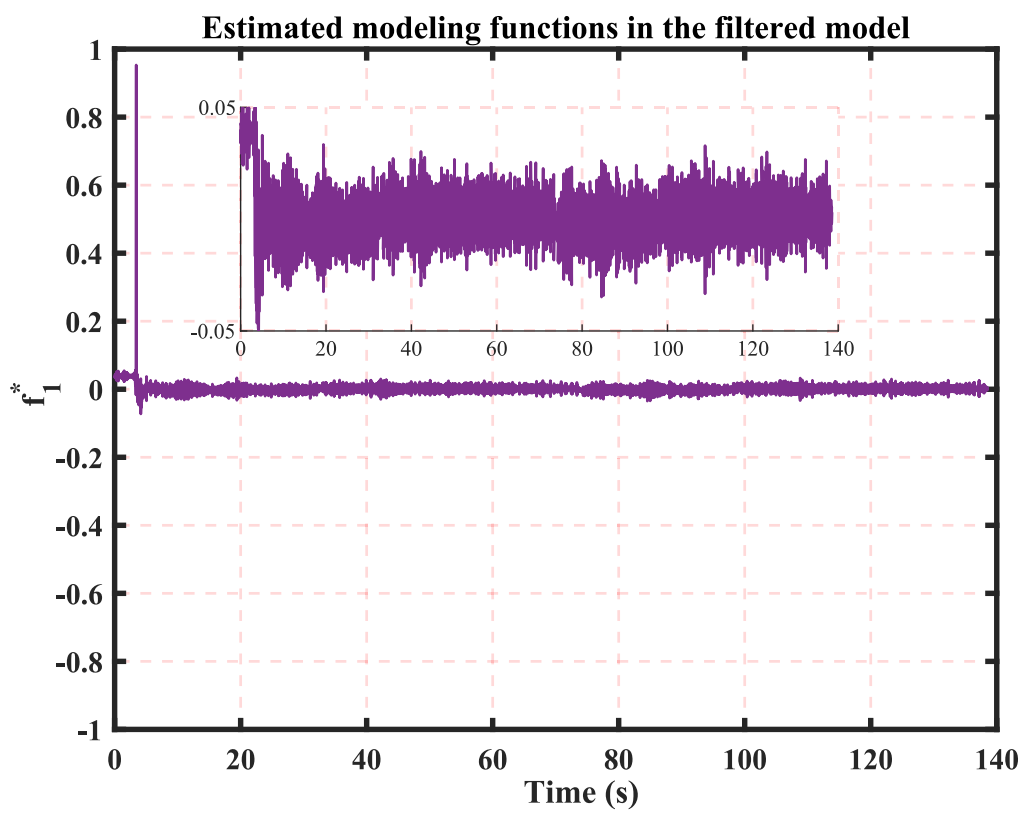

**Fig. 20.** Experiment 3: the values of $f_1^*$ (measured in N m).

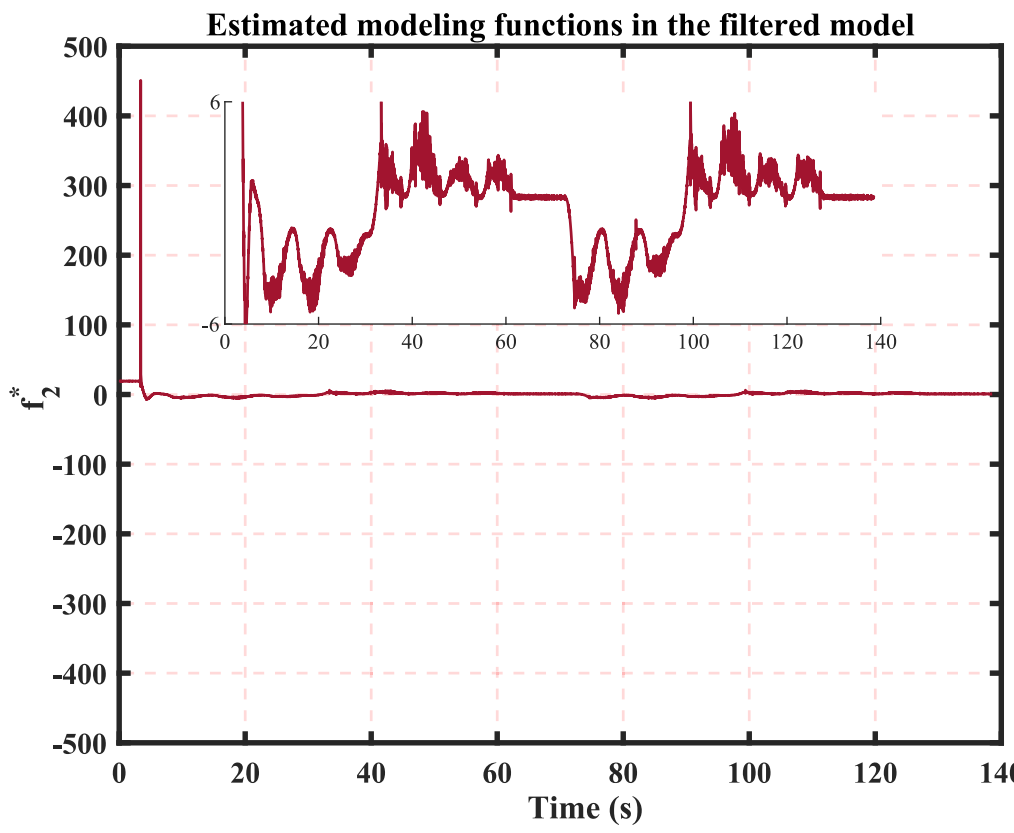

**Fig. 21.** Experiment 3: the values of $f_2^*$ (measured in N m).

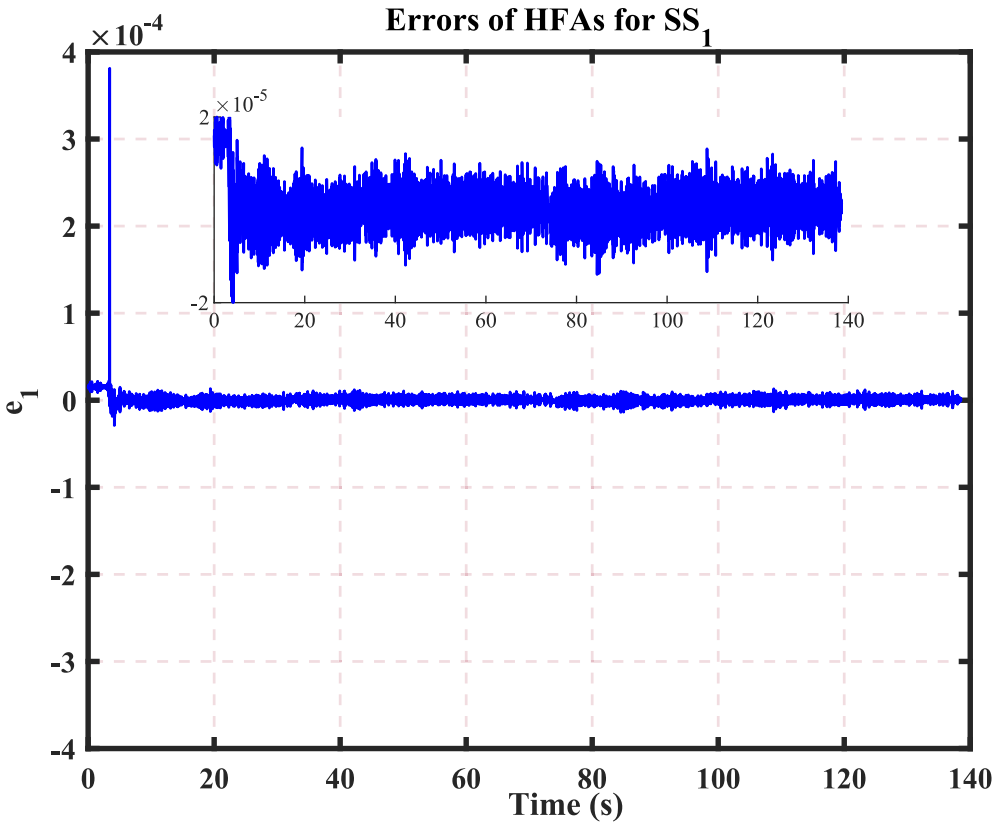

**Fig. 22.** Experiment 3: the errors between the new state of certain model in (3) estimated by employing HAEs and the state of the initial system (2) for $SS_1$.

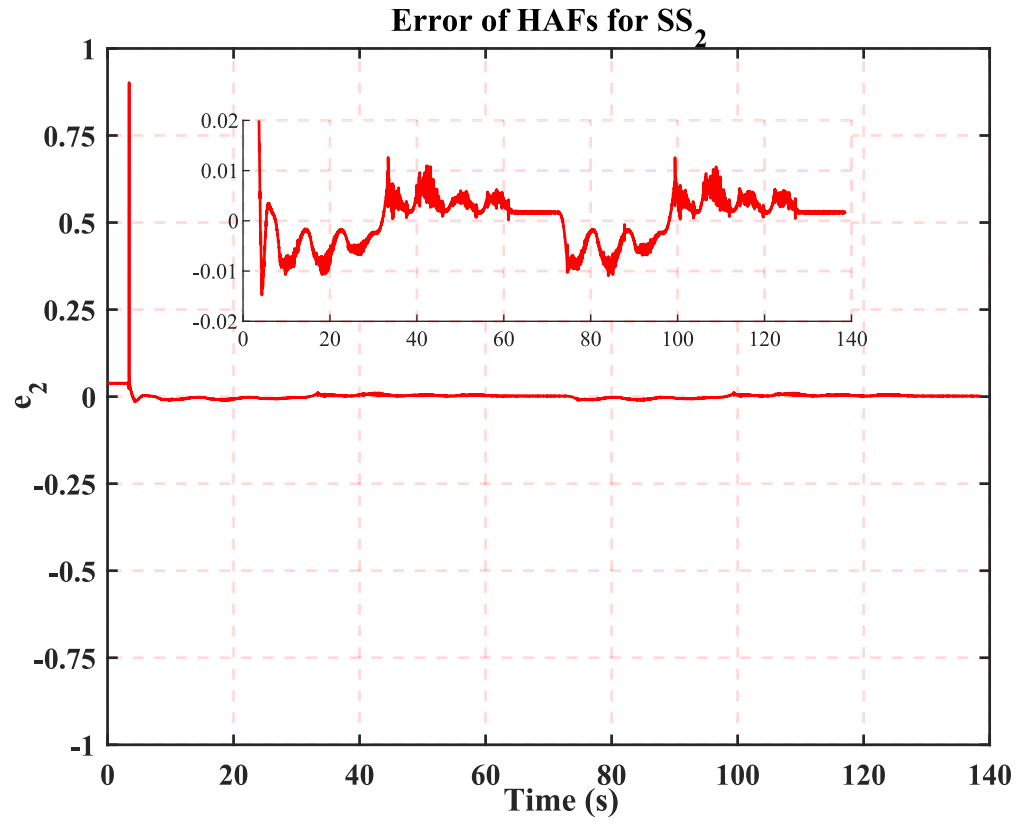

**Fig. 23.** Experiment 3: the errors between the new state of certain model in (3) estimated by employing HAEs and the state of the initial system (2) for $SS_2$.

### 5.4. Experiment 4: Trajectory tracking under 0–95% of the EMLA's capacity

In this experiment, similar to experiment 3, regardless of the system's initial point, the objective was to have it track the reference trajectory $x_{1d}$ shown in Fig. 3; however, under the high-disturbance conditions, this setup caused sudden changes similar to the disturbance imposed on the system in experiment 2. In this scenario, the controller sample time was increased, and the disturbance intensity was repeatedly varied, which could be beneficial for challenging the robustness of MRBC in tracking performance. Again, based on Section 2, and employing HAEs, the objective was to estimate $f^*_{1,2}$ and transform the uncertain system (2) into certain model system (3).

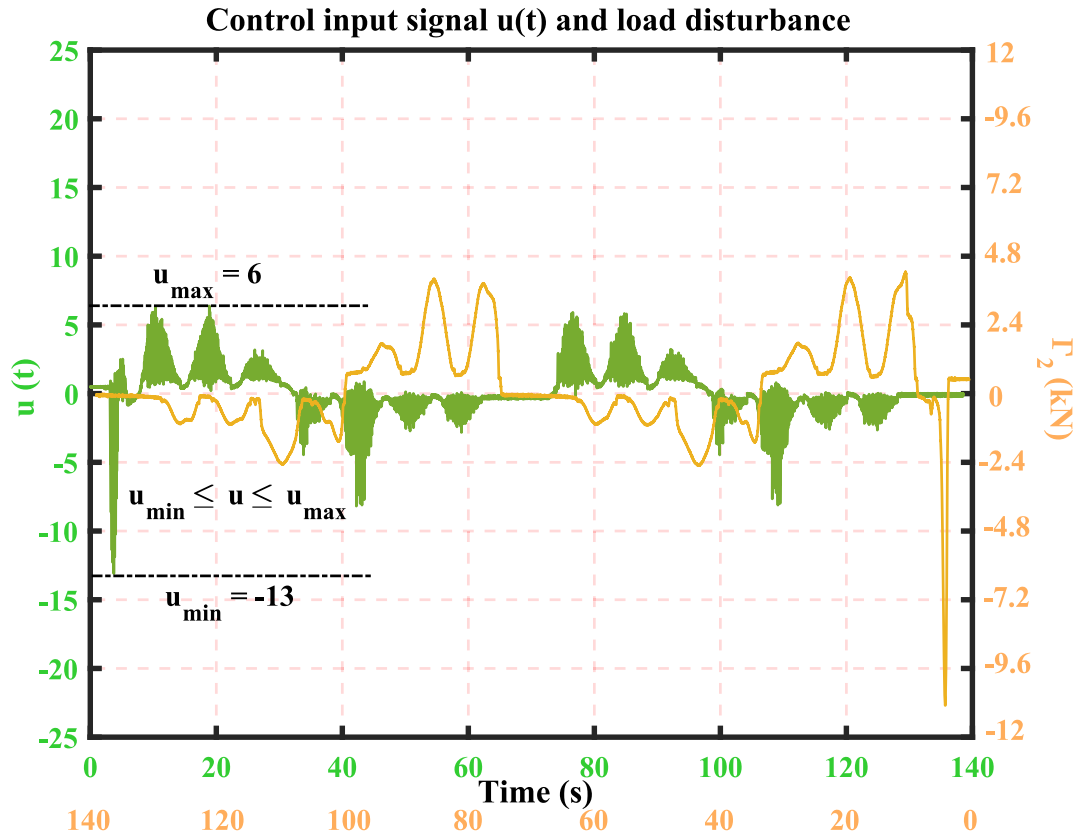

**Fig. 24.** Experiment 3: the control signal constrained by the function $\alpha(.)$ and the external force $F_L$ imposed on the system. (For interpretation of the references to color in this figure legend, the reader is referred to the web version of this article.)

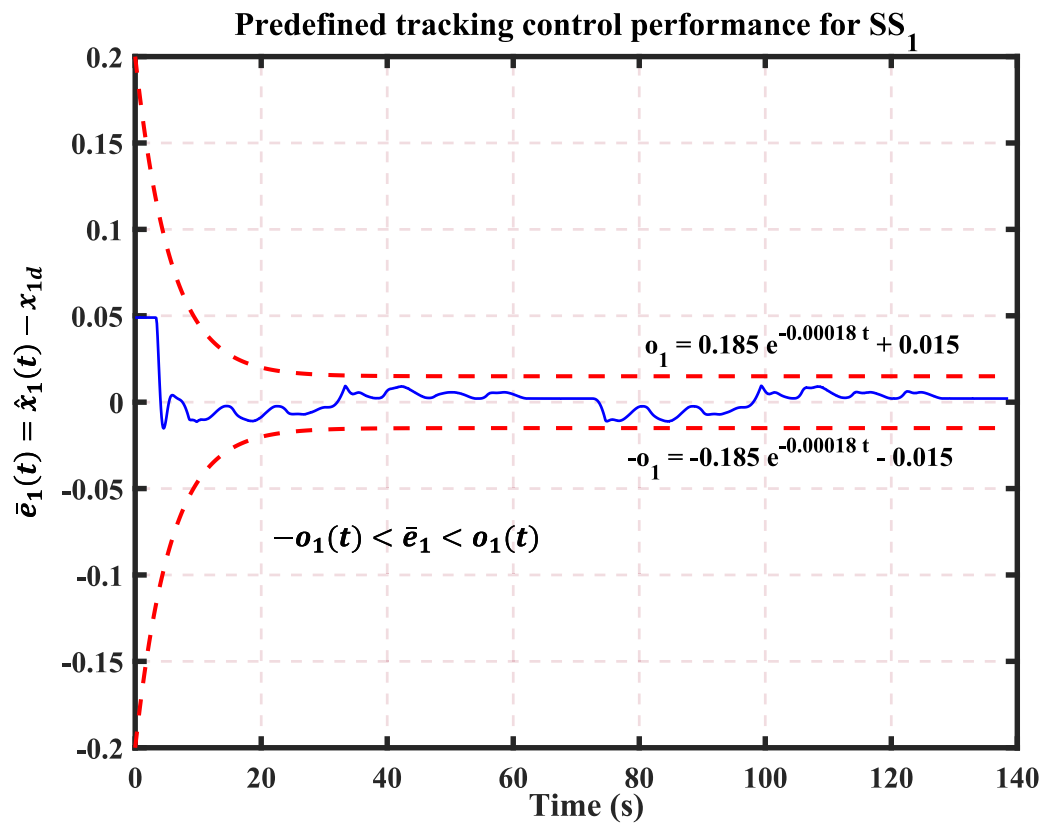

**Fig. 25.** Experiment 3: predefined control tracking performance of $\mathrm{SS}_1$.

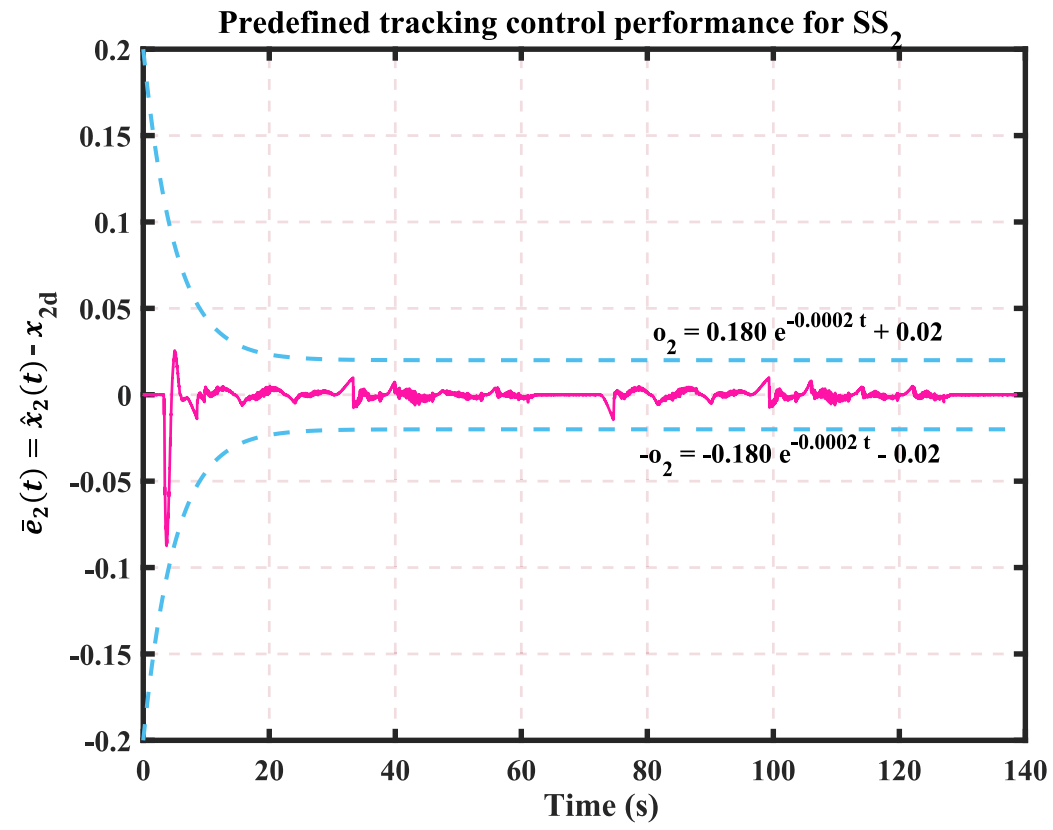

**Fig. 26.** Experiment 3: predefined control tracking performance of $\mathrm{SS}_2$.

As observed in Figs. 27 and 28, presenting severe disturbance with sudden changes in values yielded both estimated $f^*_1$ (purple trajectory) and $f^*_2$ (dark red trajectory). However, Figs. 29 and 30 show the capability of HAEs to estimate states even under varying high-disturbance conditions and track for both $\mathrm{SS}_1$ and $\mathrm{SS}_2$.

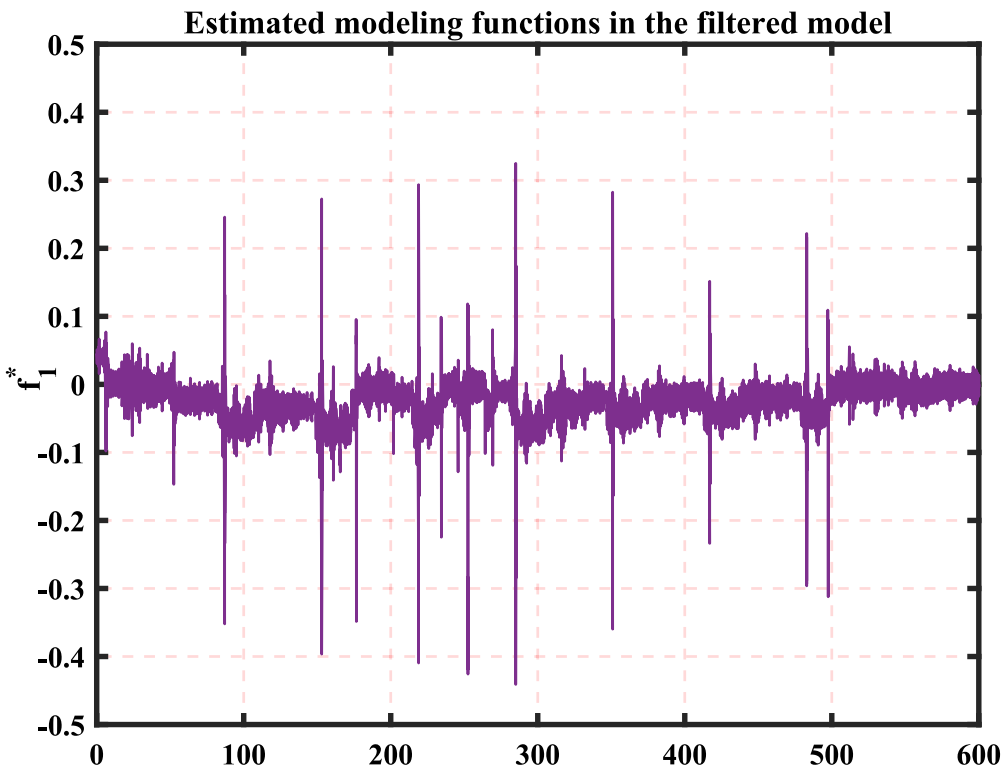

**Fig. 27.** Experiment 4: the values of $f^*_1$ (measured in N m).

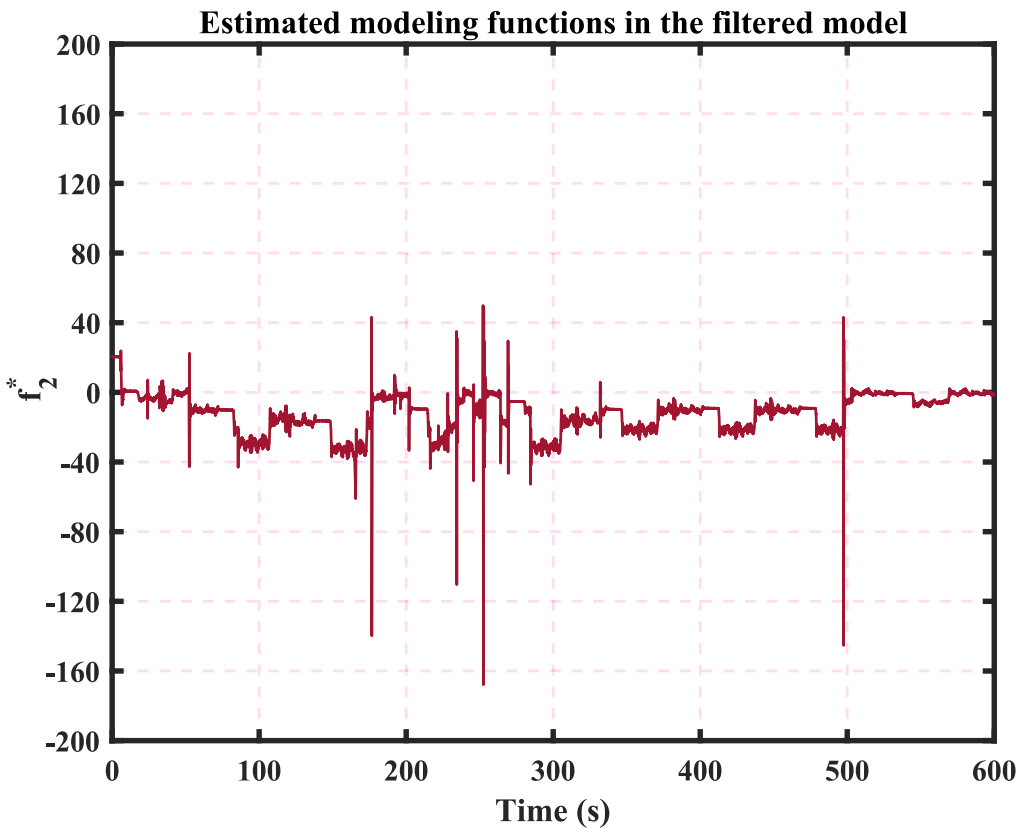

**Fig. 28.** Experiment 4: the values of $f^*_2$ (measured in N m).

The new estimated states $\hat{x}_1$ and $\hat{x}_2$ of the system were obtained separately from HAEs and input HAC-BLFs. The aim was for these states to track the reference trajectories $x_{1d}$ and $x_{2d} = \dot{x}_{1d} + \bar{f}_1 - f^*_1 = \bar{f}_1 - f^*_1$, shown Fig. 26. Similarly, the control input signal was obtained as $u = \dot{x}_{2d} + \bar{f}_2 - f^*_2$. Then, the control signal was constrained, as $-15 \leq u \leq 44.5$, following (11). Note that, as the system tracked the references under the high-disturbance condition, the values of both upper and lower bound of the control input $u_{max}$ and $u_{min}$, were considered large (near the maximum of the motor torque capacity). The green trajectory in Fig. 31 illustrates the generated control signal, which adheres to the defined constraint. The orange trajectory in this figure also shows the external force $F_L$ applied to the EMLA (extracting 0–95% of its capacity) to generate the external disturbance $\Gamma_2 = -I_{eq}^{-1} f_{eq} F_L$ in this experiment, which was about 20 times more intense than that in experiment 1, two times as intense as that in experiment 2, and four times as intense as that in experiment 3.

In this scenario, the expected control performance for $\mathrm{SS}_1$ was defined as $o_1^{shoot} = 0.2$, $o_1^{bound} = 0.015$, and $o_1^* = 0.0001$. Fig. 32 illustrates the control error for $\mathrm{SS}_1$ by employing HAC-BLFs, adhering to the PPC within $-o_1(t) < \bar{e}_1 < o_1(t)$. Similarly, the expected control performance for $\mathrm{SS}_2$ was defined as $o_2^{shoot} = 0.2$, $o_2^{bound} = 0.09$, and $o_2^* = 0.000008$. Fig. 33 illustrates the control error for $\mathrm{SS}_2$ estimated by employing HAC-BLFs, adhering to the PPC within $-o_2(t) < \bar{e}_2 < o_2(t)$.

Note that, due to sudden changes in disturbances, it was necessary to consider a wider predefined tracking performance for $\mathrm{SS}_2$ that was directly affected by the external disturbance. This is evident in Fig. 33, where an exponential convergence rate of 0.000008 was applied to the predefined performance. Table 3 presents a crucial analysis of MRBC performance, focusing on the trade-off between responsiveness and uncertainty rejection capabilities across different scenarios, as outlined in Sections 1 and 2. For clarity of analysis, data from a short period in each scenario under varying external disturbances were considered. The table indicates that as uncertainties, including disturbances and control signal saturation, become more intense, the estimation errors for states in the certain model ($e_1$ and $e_2$) increase. Note that $f_1^*$ and $f_2^*$ represent uncertainty estimations in subsystems $SS_1$ and $SS_2$ of the initial SF system and modeling terms in the reference SF system, respectively. Larger disturbances resulted in higher control effort ($\alpha(u)$) amplitudes. Similarly, tracking errors in $SS_1$ in set point-reaching or tracking scenarios improved when the control efforts were lighter, likely due to the control amplitude saturation difference $\Delta_u$ affecting the accuracy of tracking control. Interestingly, the tracking error in $SS_2$ remained similar regardless of the control efforts, although the set point-reaching scenarios demonstrated better accuracy compared to tracking scenarios, validating robustness and reliability of the proposed MRBC framework.

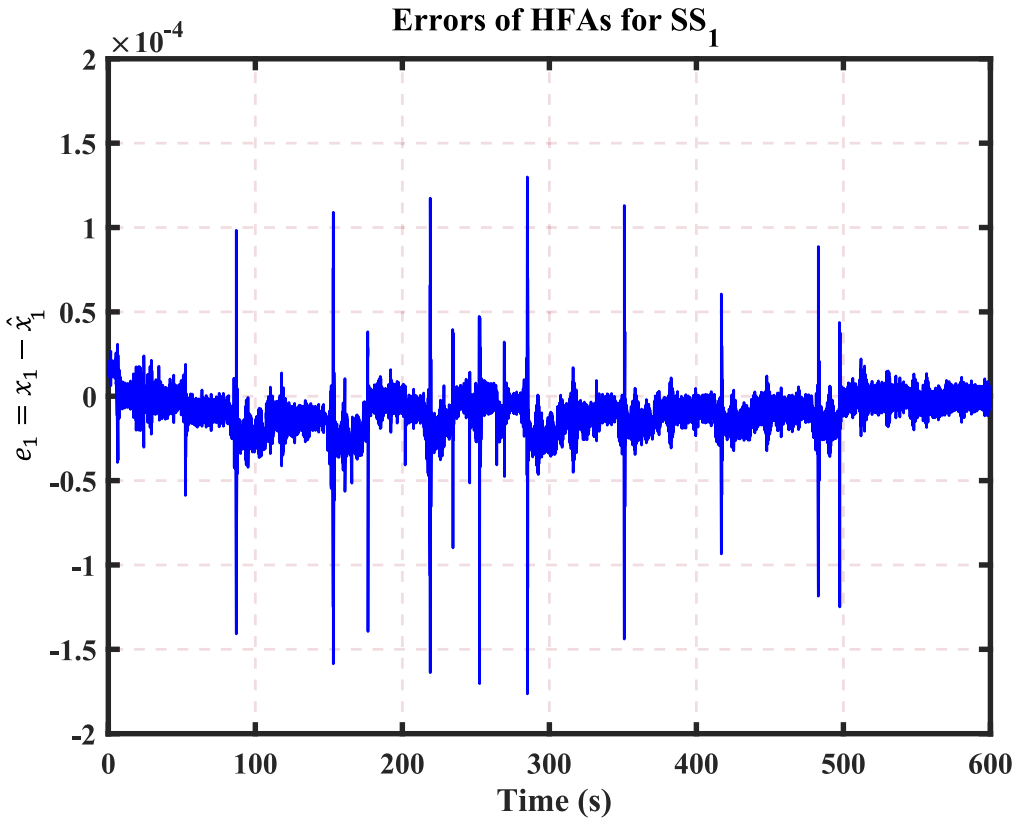

**Fig. 29.** Experiment 4: the errors between the new state of certain model in (3) estimated by employing HAEs and the state of the initial system (2) for $SS_1$.

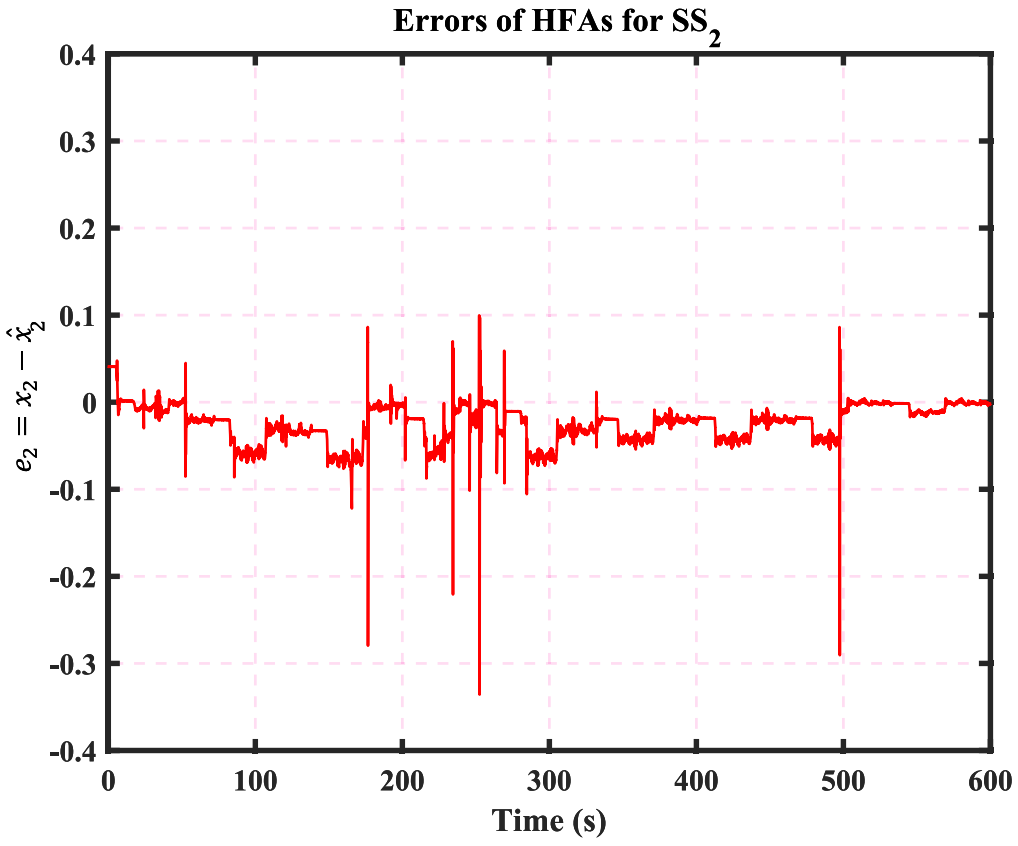

**Fig. 30.** Experiment 4: the errors between the new state of certain model in (3) estimated by employing HAEs and the state of the initial system (2) for $SS_2$.

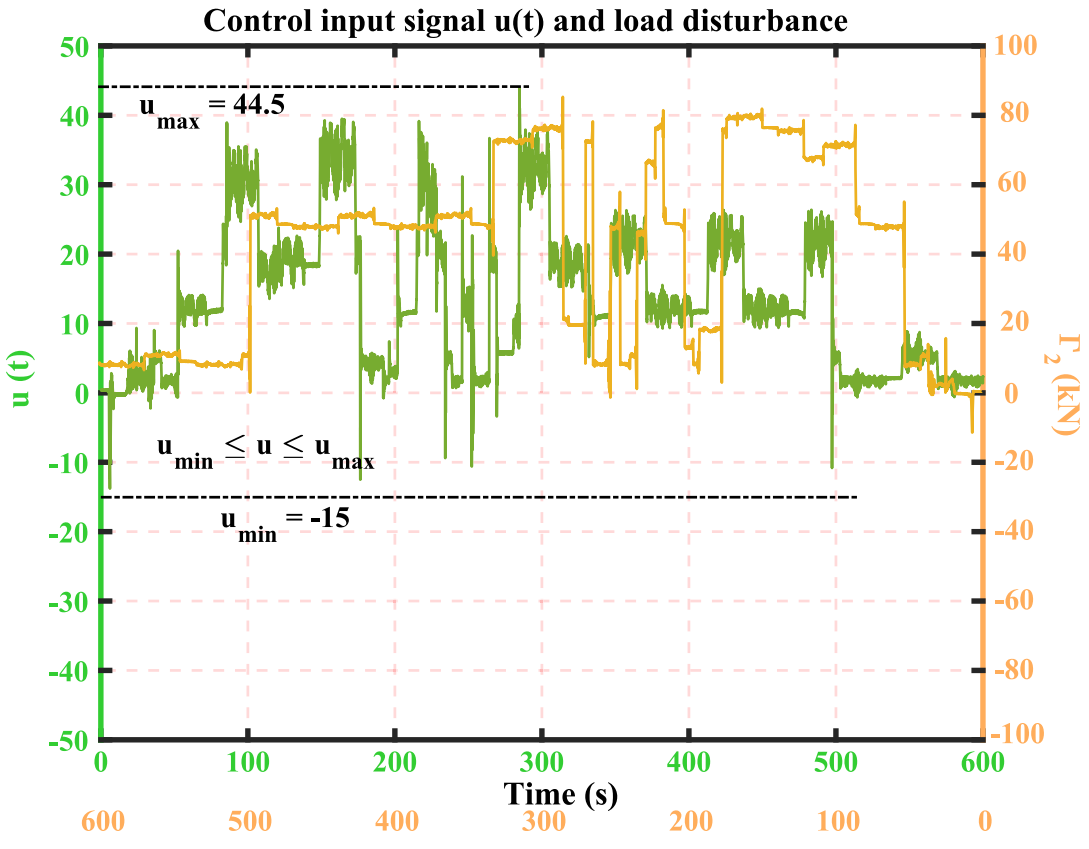

**Fig. 31.** Experiment 4: the control signal constrained by $\alpha(.)$ and the external force $F_L$ imposed on the system. (For interpretation of the references to color in this figure legend, the reader is referred to the web version of this article.)

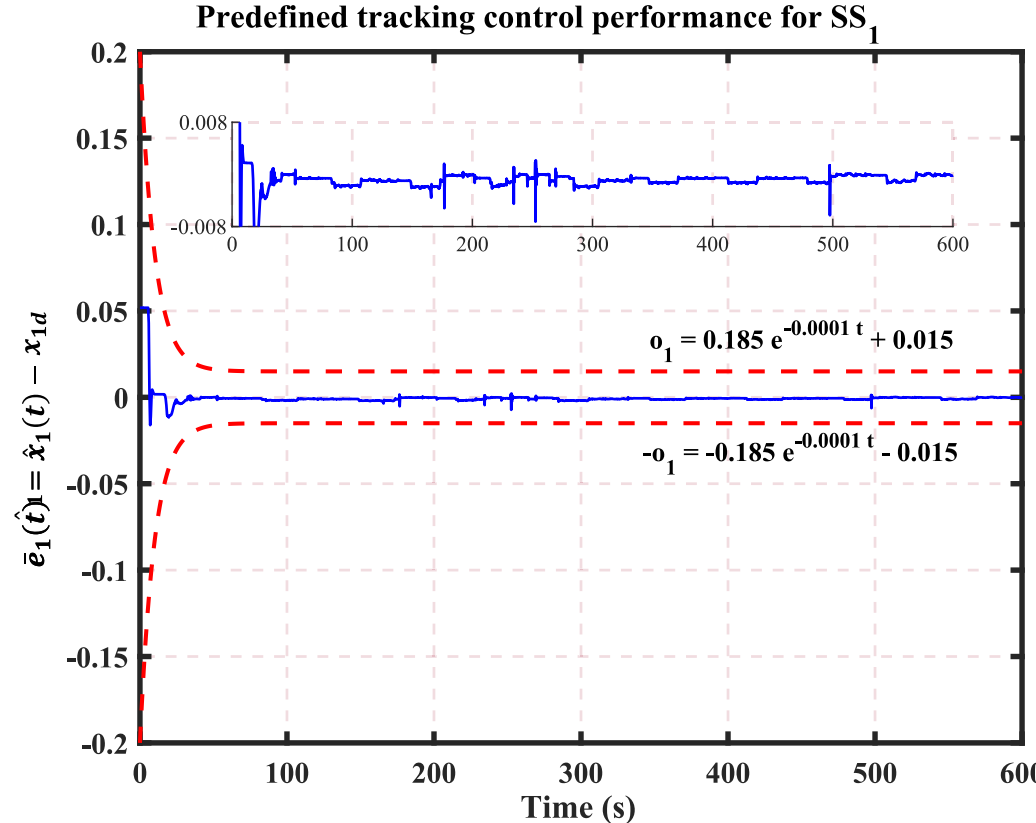

**Fig. 32.** Experiment 4: predefined control tracking performance of $SS_1$.

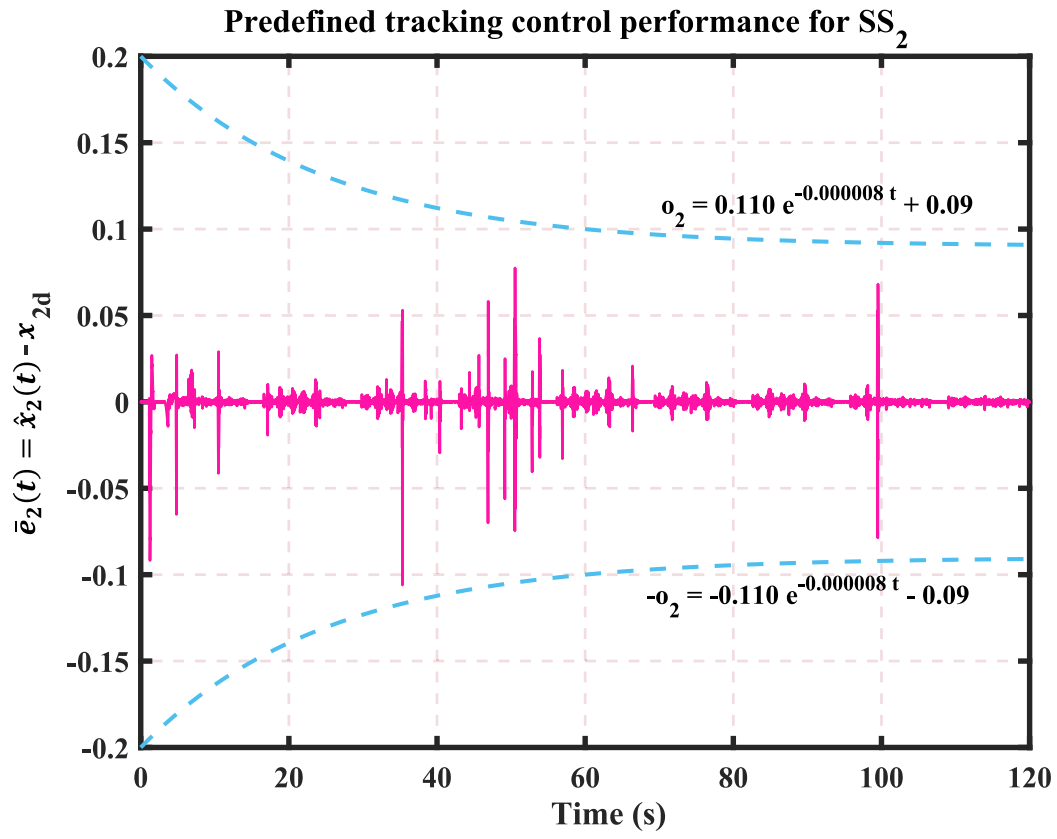

**Fig. 33.** Experiment 4: predefined control tracking performance of $SS_2$.

### 5.5. Experiment 5: comparative results under 60%–85% of the EMLA's capacity

In this scenario, the proposed MRBC framework is compared against two state-of-the-art high-performance adaptive control strategies: the AATC proposed in Zhao et al. (2021), and the AFTC proposed in Guo and Zhang (2024). Both methods were designed to achieve PPC and demonstrated strong performance in similar applications. Therefore, they were selected to provide a valid and meaningful basis for comparison. All three control strategies – the proposed MRBC, AATC, and AFTC – were implemented on the studied EMLA to track a repetitive trapezoidal reference profile, tackling the presence of parameter uncertainty in the load. In this scenario, the EMLA was commanded to hold its initial position at 5 cm for 5 s. It then moved forward with a gradually varying velocity—accelerating up to 5 cm/s and then decelerating back to zero. Upon reaching the target position of 25 cm, it held that position for 10 s. The system then returned to the initial position following the same velocity profile in reverse. This cycle was repeated periodically.

**Table 3**
MRBC performance analysis in terms of responsiveness and rejecting uncertainty capability.

| Performance criteria | $\lvert F_L \rvert \approx 1$ Experiment 1 ($30 \le t \le 50$) | $\lvert F_L \rvert \approx 38$ Experiment 2 ($80 \le t \le 90$) | $\lvert F_L \rvert \approx 9$ Experiment 3 ($85 \le t \le 90$) | $\lvert F_L \rvert \approx 80$ Experiment 4 ($150 \le t \le 175$) |
|---|---|---|---|---|
| $\sup \lvert f_1^* \rvert$ | $1 \times 10^{-2}$ | $6 \times 10^{-2}$ | $4 \times 10^{-2}$ | $9 \times 10^{-1}$ |
| $\sup \lvert f_2^* \rvert$ | 1 | 26 | 5.5 | 39 |
| $\sup \lvert e_1 \rvert$ | $1 \times 10^{-5}$ | $2.7 \times 10^{-5}$ | $1.8 \times 10^{-5}$ | $6 \times 10^{-5}$ |
| $\sup \lvert e_2 \rvert$ | $1.8 \times 10^{-3}$ | $5.2 \times 10^{-2}$ | $1 \times 10^{-2}$ | $6 \times 10^{-2}$ |
| $\sup \lvert \alpha(u) \rvert$ | 0.4 | 28 | 6 | 40 |
| $\sup \lvert \bar{e}_1 \rvert$ | $1.1 \times 10^{-4}$ | $1.5 \times 10^{-3}$ | $1.4 \times 10^{-3}$ | $1.1 \times 10^{-3}$ |
| $\sup \lvert \bar{e}_2 \rvert$ | $1 \times 10^{-4}$ | $1 \times 10^{-4}$ | $2.5 \times 10^{-4}$ | $2.5 \times 10^{-4}$ |

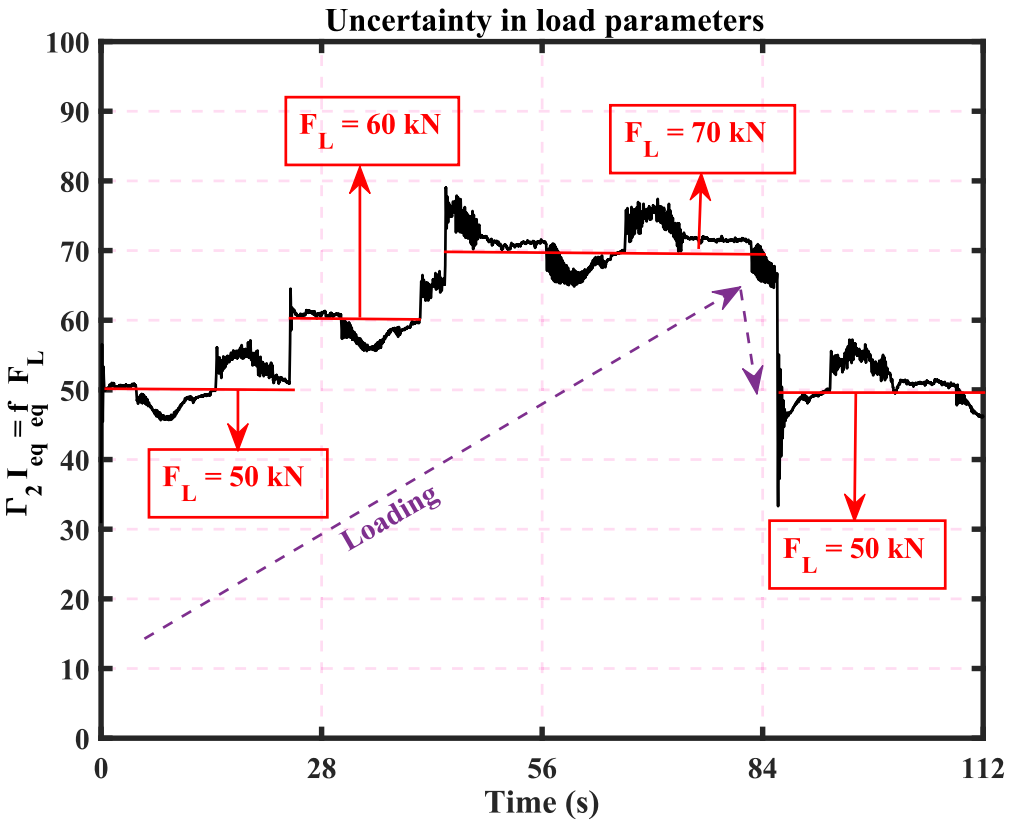

**Fig. 34.** Experiment 5: external force with parameter uncertainties.

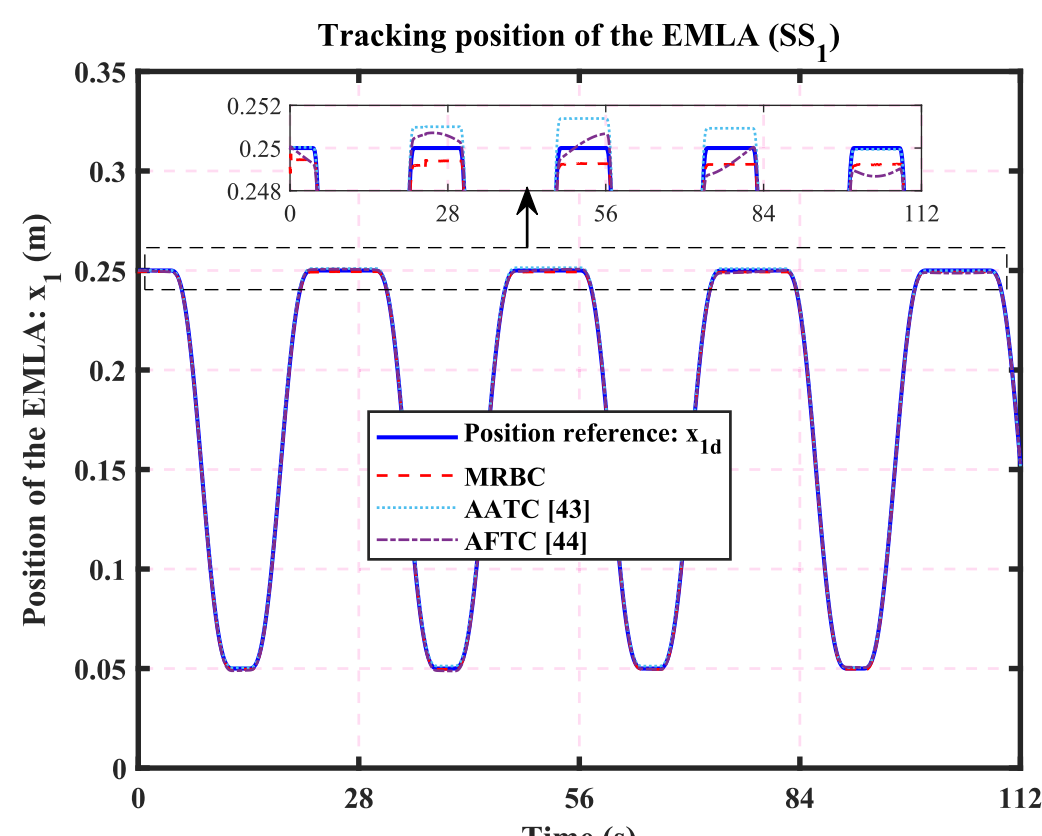

**Fig. 35.** Experiment 5: tracking control for $SS_1$.

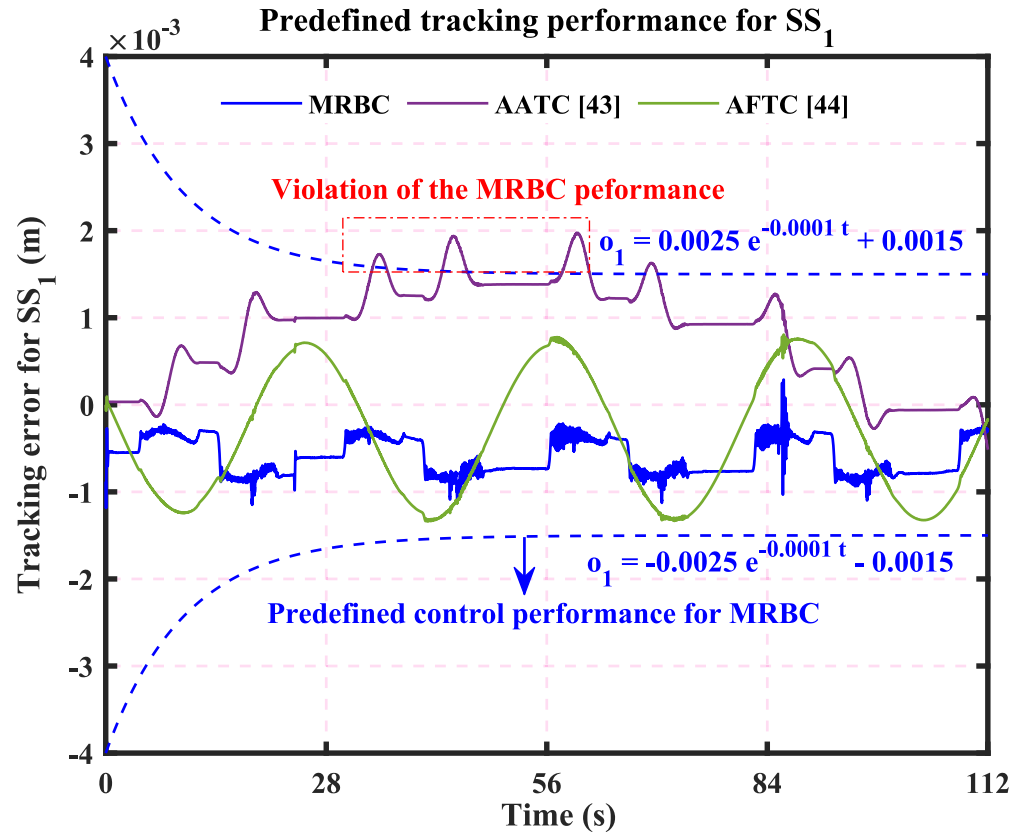

**Fig. 36.** Experiment 5: comparative tracking error of $SS_1$.

To ensure a fair comparison, Eq. (26) was employed to determine the suboptimal control gains, while also ensuring that the EMLA operation started from the same initial point in each experiment. Fig. 34 illustrates the time evolution of the load modeling term $\Gamma_2 I_{eq} = f_{eq} F_L$ in kN ($\Gamma_2$ with unit N m is the external disturbance of the EMLA; see (25)), representing the pushing force generated by the hydraulic cylinder coupled with the EMLA system. To introduce parametric uncertainty into the control system, the equivalent inertia $I_{eq}$ and equivalent damping $f_{eq}$ were assumed to be unknown in the control frameworks. As a result, only the external load force $F_L$ could be adjusted directly during experimentation. As shown in the figure, $F_L$ was varied step-wise across four intervals to simulate load uncertainties: starting at 50 kN, then increased to 60 kN and 70 kN, followed by a return to 50 kN. The corresponding values reflect these changes, showing step transitions and transient fluctuations due to the unmodeled dynamics and nonlinearities. The dashed purple arrow labeled "Loading" indicates the progressive increase and subsequent reduction in load, which challenges the control system's ability to adapt to significant and abrupt variations in load dynamics.

Fig. 35 shows the tracking performance of the EMLA system position state $x_1$ under the proposed MRBC controller, compared with AATC (Zhao et al., 2021) and AFTC (Guo & Zhang, 2024), during the experiment. The controllers were evaluated while tracking the repetitive trapezoidal reference trajectory $x_{1d}$, under conditions of parameter uncertainty and varying external loads. The EMLA was commanded to stop and hold position at specific points-namely 0.05 m and 0.25 m -as part of the reference profile. While all three controllers demonstrated effective tracking performance throughout the motion phases, slight steady-state deviations can be observed at the intended stopping positions (0.05 m and 0.25 m) as highlighted in the zoomed inset. These minor discrepancies reflect the inherent challenge of maintaining precise position regulation in the presence of significant load-induced disturbances and parametric uncertainties in the system model.

Following Fig. 35, the comparative position tracking error performance for $SS_1$ in the EMLA is presented in Fig. 36, using the three control strategies. The objective was to evaluate the controllers under a strict PPC, as illustrated by the symmetric time-varying bounds (shown in dashed blue curves). This figure is also validated that all controllers achieved high tracking accuracy, maintaining absolute position errors below 2 mm throughout the entire test duration. However, under the applied PPC evaluation, the AATC controller slightly and repeatedly violated the performance bounds, particularly during periods of high load forces at 60 kN and 70 kN (see Fig. 34). In contrast, both the MRBC and AFTC controllers successfully adhered to the predefined performance limits, with the proposed MRBC approach demonstrating the best overall performance. It maintained tighter error bounds, exhibited faster convergence, and showed minimal deviation from the desired trajectory within the constraint boundaries.

Following position tracking, the tracking performance of the velocity state $x_2$ (in $SS_2$) of the EMLA system during this experiment is shown in Figs. 37 and 38. As observed, the system was commanded to follow the repetitive velocity reference $x_{2d}$, while being subjected to significant load variations. The velocity tracking response implicitly demonstrated the robustness of the controllers to external load forces with parametric uncertainties, particularly due to unmodeled variations in inertia and damping. All three controllers were able to follow the desired velocity trajectory with high accuracy; however, differences in smoothness and transient response are evident. The most significant impact of load disturbances was observed during the sudden load reduction from 70 kN to 50 kN at approximately 84 s. This event introduced a noticeable dynamic perturbation, challenging the controllers' robustness and disturbance rejection capabilities. As shown in the figure, this load transition affected the tracking accuracy and transient behavior, particularly in the velocity response. The proposed MRBC controller exhibited smoother velocity tracking, particularly during load transitions, as shown in the zoomed-in inset. It produced a less oscillatory and more stable response compared to AATC and AFTC, indicating better disturbance rejection and dynamic adaptation.

**Table 4**
Comparison of the MRBC performance with AATC (Zhao et al., 2021) and AFTC (Guo & Zhang, 2024) in terms of accuracy, response speed, and robustness under parametric uncertainties.

| Performance criteria | MRBC in Experiment 5 | AATC (Zhao et al., 2021) in Experiment 5 | AFTC (Guo & Zhang, 2024) in Experiment 5 |
|---|---|---|---|
| Accuracy in $SS_1$ (m) | $5 \times 10^{-4}$ | $11 \times 10^{-4}$ | $8 \times 10^{-4}$ |
| Accuracy in $SS_2$ (m/s) | $1 \times 10^{-3}$ | $5 \times 10^{-3}$ | $4 \times 10^{-3}$ |
| Transient time in $SS_1$ (s) | 0.42 | 0.85 | 0.71 |
| Transient time in $SS_2$ (s) | 0.43 | 0.72 | 0.65 |
| Max. control effort (N m) | 33.1 | 33.0 | 36.5 |
| Robustness (50–70 kN load) | Met all constraints | Violated position constraints | Met all constraints |
| Robustness (70–50 kN load) | Met all constraints | Met all constraints | Violated torque constraints |

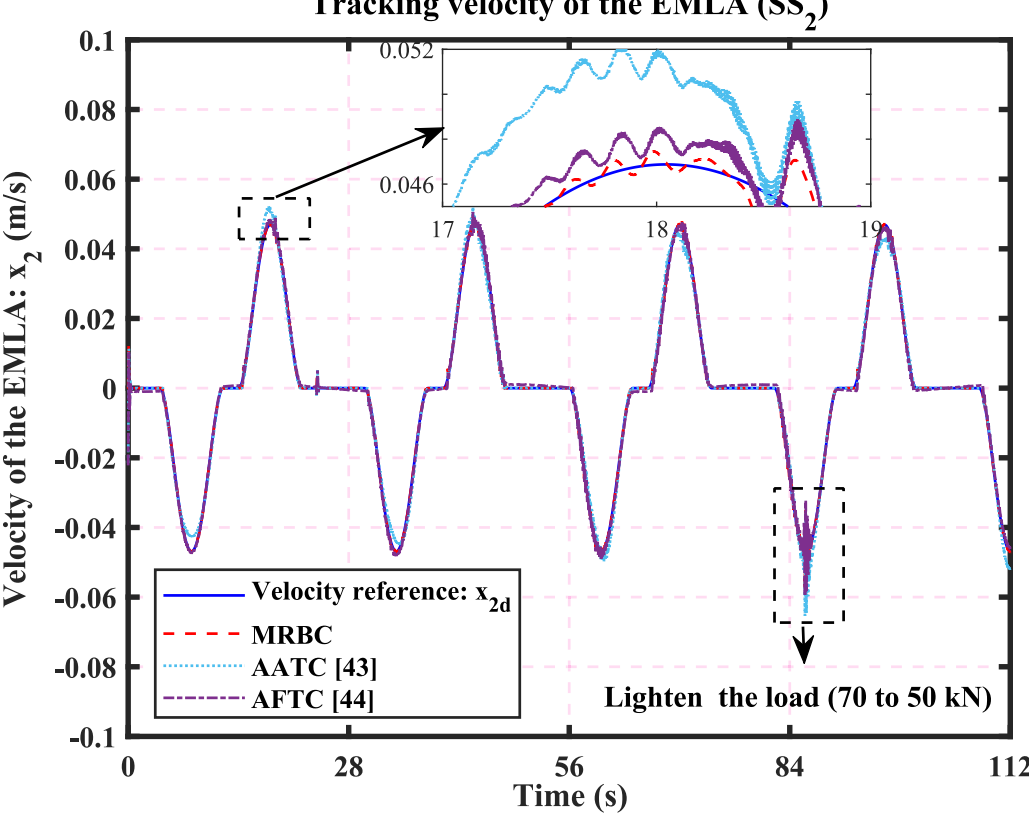

**Fig. 37.** Experiment 5: tracking control for $SS_2$.

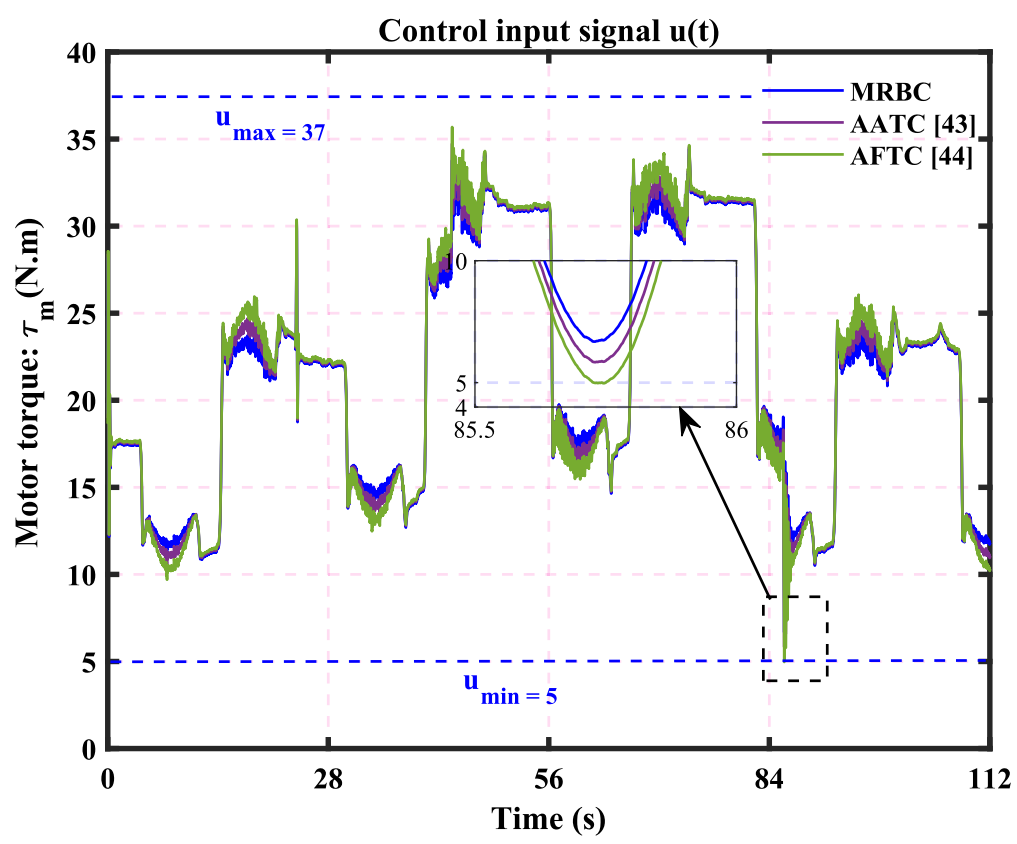

**Fig. 39.** Experiment 5: Comparative control inputs constrained by $\alpha(.)$.

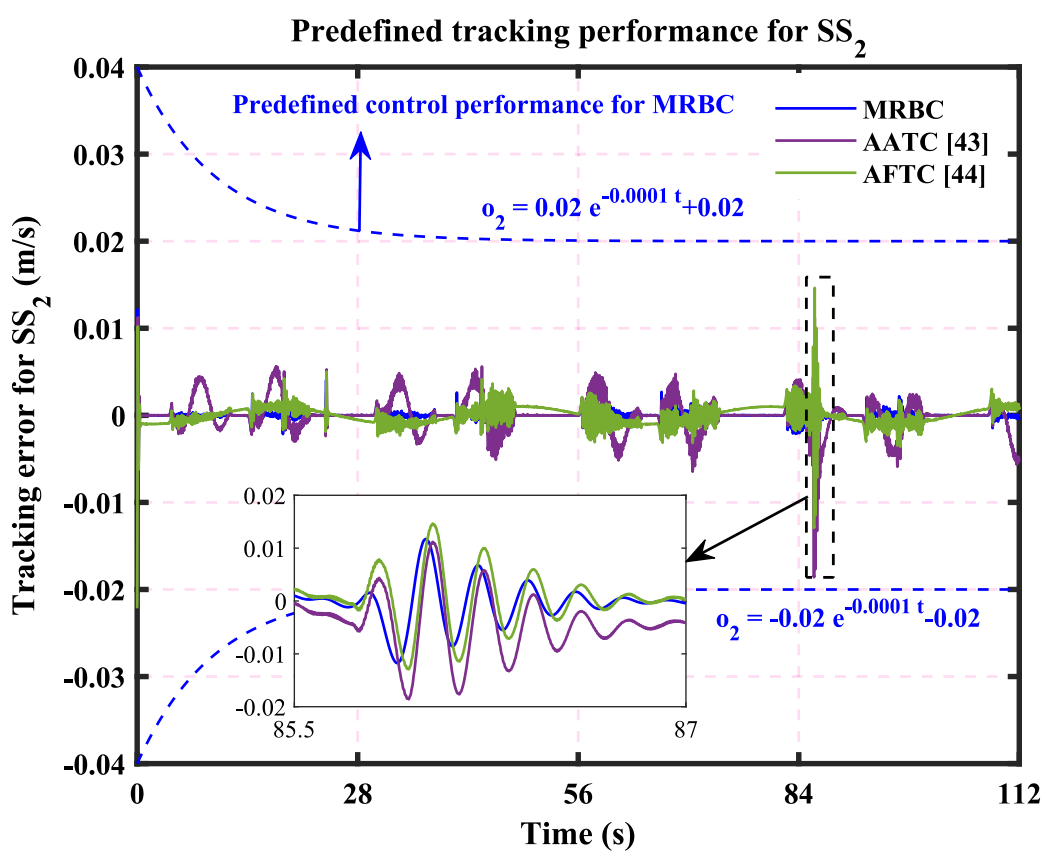

**Fig. 38.** Experiment 5: Comparative tracking error of $SS_2$.

In Fig. 38, the tracking errors for the MRBC, AATC, and AFTC controllers were also compared against the predefined control performance bounds defined by a time-varying PPC (shown as dashed blue curves). All three control strategies successfully adhered to the PPC constraint, maintaining the tracking error strictly within the upper and lower bounds $\pm 0.02e^{-0.0001t} + 0.02$ m/s throughout the experiment. The zoomed-in region highlights the controller behavior during the sudden load reduction from 70 kN to 50 kN at ~84 s, where transient error responses were most prominent. Among the three, the proposed MRBC controller demonstrated smoother error dynamics with lower oscillation amplitudes and faster convergence, confirming its improved robustness and precision under parametric uncertainty and external disturbances. AATC and AFTC also showed good performance but exhibited slightly more pronounced oscillations in certain intervals. Finally, Fig. 39 presents the control input signals, specifically the motor torque $\tau_m$, generated by the MRBC, AATC, and AFTC controllers. These signals correspond to the torque applied by the PMSM to drive the EMLA under varying hydraulic load conditions. The control inputs were constrained by predefined upper and lower bounds, $u_{\min} = 5$ Nm and $u_{\max} = 37$ Nm, as enforced by the saturation function $\alpha(\cdot)$. All three controllers generated feasible torque profiles to follow the reference motion trajectory. However, notable differences in control effort and constraint adherence are evident. The MRBC and AATC controllers demonstrated lower and smoother control effort, successfully remaining within the defined torque bounds throughout the entire operation. In contrast, the AFTC controller reached the torque constraint during the abrupt load reduction from 70 kN to 50 kN at around 84 s, as highlighted in the zoomed inset. During this event, the saturation function $\alpha(\cdot)$ was triggered to enforce the torque limit and prevent actuator overdrive. All these results indicate that the proposed MRBC framework not only ensured effective reference tracking but also achieved better control performance and compliance with input constraints, even under significant load variations and parametric uncertainty.

In summary, Table 4 presents the comparative evaluation of the proposed MRBC controller against two other benchmark strategies, AATC (Zhao et al., 2021) and AFTC (Guo & Zhang, 2024) in the same condition, based on accuracy, response speed, control effort, and robustness for the two EMLA subsystems: $SS_1$ (position tracking) and $SS_2$ (velocity tracking), under parameter uncertainties and varying external loads. The entries highlighted in red indicate the lowest performance among the three controllers during the operation in each metric. It shows that the MRBC controller achieved the highest accuracy (mean-squared error), with the lowest tracking errors in both subsystems: $5 \times 10^{-4}$ m for $SS_1$ and $1 \times 10^{-3}$ m/s for $SS_2$. The AATC controller, marked in red, showed relatively higher tracking errors, particularly in $SS_2$, indicating a slightly reduced effectiveness in velocity regulation compared to MRBC. The AFTC controller exhibited moderately better accuracy than AATC but still did not reach the precision level achieved by MRBC. Similarly, MRBC provided the fastest transient response, with settling times of 0.42 s $(SS_1)$ and 0.43 s $(SS_2)$. In contrast, AATC again showed the slowest response times, marked in red, reflecting its delayed convergence under dynamic conditions. AFTC responded faster than AATC but slower than MRBC. In addition, the amplitude of the generated torque (control effort) by MRBC and AATC was lower than that of AFTC (highlighted in red at 36.5 Nm) during the operation. Finally, the MRBC satisfied all performance constraints under both load-increasing (50–70 kN) and load-decreasing (70–50 kN) scenarios. However, AATC violated position constraints under high-load conditions, while AFTC violated torque constraints during load reduction-both cases marked in red. While both AATC and AFTC demonstrated robustness under varying load conditions, their performance was less consistent compared to MRBC, particularly in maintaining all prescribed constraints. Indeed, both AATC and AFTC exhibited control performance close to that of MRBC, but it is important to note that the EMLA system under study serves as a prototype for one of the joints in an electrically actuated heavy-duty manipulator, which will eventually include two additional EMLA-actuated joints. In such multi-joint systems, even minor tracking deviations – on the order of millimeters – in a single actuator can accumulate and result in significant positioning errors at the manipulator's end-effector, particularly when executing complex tasks. Therefore, the better precision and robustness of the proposed MRBC controller make it a more suitable and reliable option for this class of high-performance robotic applications.

## 6. Conclusion

This paper proposed theoretical foundations, which were validated in practice, for controlling a class of $n$-order SF systems including parametric and modeling uncertainties, unknown control gain functions, external disturbances, and control signal amplitude saturation effects. Auxiliary tools, along with a specific stability connector, were presented to address dynamic interactions among neighboring SSs, enabling uniformly exponentially stable control while avoiding excessive growth in control design complexity and ensuring tracking responses remain within user-defined performance criteria. The proposed method included HAEs to convert an uncertain SF system into a certain reference model and HAC-BLFs to guarantee tracking responses within a specified transient and steady-state response. Theoretical developments on uniformly exponential convergence (in Theorem 4.1) and adhering to the predefined tracking control performance were verified in experimental tests conducted on an EMLA with uncertain SF form subjected to external disturbances varying up to 95% of the EMLA's capacity. The PMSM-powered EMLA studied in Section 5 is a single-degree-of-freedom (1-DoF) linear actuator, serving as a prototype for actuating one joint of a fully electrified 3-DoF heavy-duty manipulator currently under development. Therefore, future research will focus on extending the MRBC framework to multi-input multi-output (MIMO) SF systems, corresponding to the full 3-DoF manipulator actuated by three interdependent PMSM-driven EMLAs, which exhibit mutual dynamic interactions and are intended to operate collaboratively within the manipulator's task space.

## CRediT authorship contribution statement

**Mehdi Heydari Shahna:** Writing – original draft, Visualization, Validation, Software, Methodology, Investigation, Formal analysis, Data curation, Conceptualization. **Jukka-Pekka Humaloja:** Writing – review & editing, Supervision. **Jouni Mattila:** Writing – review & editing, Supervision, Resources, Funding acquisition.

## Declaration of competing interest

The authors declare that they have no known competing financial interests or personal relationships that could have appeared to influence the work reported in this paper.

## Acknowledgments

This work was supported by the Business Finland Partnership Project, 'Future All-Electric Rough Terrain Autonomous Mobile Manipulators' under Grant No. 2334/31/2022.

## Appendix A. Proof of Theorem 2.1

A quadratic function for $SS_1$ is suggested as follows:

$$V_1 = \frac{1}{2}\ [e_1^2 + \tilde{\Psi}_1^2] \tag{28}$$

From $\dot{e}_1$ and $\dot{\tilde{\Psi}}_1$ definitions in (6) and (5) and knowing $e_0 = 0$:

$$\dot{V}_1 = e_1 e_2 + e_1 d_1^* + e_1 \Gamma_1 - \xi_1 \tilde{\Psi}_1 e_1^2 - \frac{1}{2}\lambda_1 e_1^2 - \beta_1 \tilde{\Psi}_1^2 + \xi_1 e_1^2 \tilde{\Psi}_1 \tag{29}$$

From Assumption 2.1 and using Young's inequality and arbitrary positive constants $\delta_1$ and $\zeta_1$:

$$\dot{V}_1 \le e_1 e_2 + \frac{\delta_1}{2} e_1^2 + \frac{1}{2\delta_1} d_1^{*2} + \frac{\zeta_1}{2} e_1^2 + \frac{1}{2\zeta_1}\Gamma_1^2 - \frac{1}{2}\lambda_1 e_1^2 - \beta_1 \tilde{\Psi}_1^2 \tag{30}$$

Therefore,

$$\dot{V}_1 \le e_1 e_2 - \frac{\lambda_1 - \delta_1 - \zeta_1}{2} e_1^2 - \beta_1 \tilde{\Psi}_1^2 + \frac{1}{2\delta_1} d_1^{*2} + \frac{1}{2\zeta_1}\Gamma_1^2 \tag{31}$$

Let us define:

$$\begin{aligned} \rho_1 &= \min[\lambda_1 - \delta_1 - \zeta_1,\ 2\beta_1], \\ \ell_1 &= \frac{1}{2\delta_1} d_1^{*2} + \frac{1}{2\zeta_1}\Gamma_1^2 \end{aligned} \tag{32}$$

To guarantee $\rho_1 > 0$, $\lambda_1$ should be selected large enough to satisfy $\lambda_1 > \delta_1 + \zeta_1$. Hence, from (8), (32), and (31),

$$\dot{V}_1 \le s_1 - \rho_1 V_1 + \ell_1 \tag{33}$$

Similarly, quadratic functions are defined for $SS_i$ for $i = 2, \dots, n-1$, as

$$V_i = \frac{1}{2}\ [e_i^2 + \tilde{\Psi}_i^2] \tag{34}$$

and reaching

$$\dot{V}_i \le e_i e_{i+1} - e_i e_{i-1} - \rho_i V_i + \ell_i = s_i - s_{i-1} - \rho_i V_i - \rho_i V_i + \ell_i \tag{35}$$

where

$$\begin{aligned} \rho_i &= \min[\lambda_i - \delta_i - \zeta_i,\ 2\beta_i], \\ \ell_i &= \frac{1}{2\delta_i} d_i^{*2} + \frac{1}{2\zeta_i}\Gamma_i^2 \end{aligned} \tag{36}$$

Finally, a similar quadratic function is defined for $SS_n$, as

$$V_n = \frac{1}{2}\ [e_n^2 + \tilde{\Psi}_n^2] \tag{37}$$

and reaching

$$\dot{V}_n \le -e_n e_{n-1} - \rho_n V_n + \ell_n = -s_{n-1} - \rho_n V_n + \ell_n \tag{38}$$

where

$$\rho_n = \min[\lambda_n - \delta_n - \zeta_n,\ 2\beta_n], \\ \ell_n = \frac{1}{2\delta_n} d_n^{*2} + \frac{1}{2\zeta_n} \Gamma_n^2 \tag{39}$$

Now, a quadratic function is defined for the entire SF system employing HAEs:

$$V_{\text{ob}} = V_1 + \sum_{i=2}^{n-1} V_i + V_n = \frac{1}{2}(\boldsymbol{e}_{ob}^{\top}\boldsymbol{e}_{ob} + \boldsymbol{\Psi}^{\top}\boldsymbol{\Psi}) \tag{40}$$

where $\boldsymbol{e}_{ob} = \left[e_1, \ldots, e_n\right]^T$ and $\boldsymbol{\Psi} = \left[\tilde{\Psi}_1, \ldots, \tilde{\Psi}_n\right]^T$. The derivative of (40), based on (33), (35), and (38), can be expressed as

$$\dot{V}_{ob} \leq \sum_{i=1}^{n-1} \left[-s_i + s_i\right] - \rho_{ob} V_{ob} + \ell_{ob} \tag{41}$$

where:

$$\rho_{ob} = \min[\rho_1, \ldots, \rho_n] \\ = \min[\min_{i=1}^{n} \left\{\lambda_i - \delta_i - \zeta_i,\ 2\beta_i\right\}] \\ \ell_{ob} = \ell_1 + \cdots + \ell_n = \sum_{i=1}^{n} \frac{1}{2\delta_i} d_i^{*2} + \frac{1}{2\zeta_i} \Gamma_i^2 \tag{42}$$

As observed, term $e_{i+1}$ for $SS_i$ was treated as an external input that caused $s_i$ to appear in the $i$th SS. Hence, using the proposed HAEs (4), instability terms $s_i$ in each SS, provided in (8), were canceled out for the entire system's stability. Thus:

$$\dot{V}_{ob} \leq -\rho_{ob} V_{ob} + \ell_{ob} \tag{43}$$

(43) can be solved as follows:

$$V_{ob} \leq V_{ob}\left(t_0\right) e^{-\left\{\rho_{ob}\left(t-t_0\right)\right\}} + \ell_{ob} \int_{t_0}^{t} e^{\left\{-\rho_{ob}(t-T)\right\}} \quad dT \tag{44}$$

Because $e^{-\rho_{ob}(t-t_0)}$ is always decreasing,

$$V_{ob} \leq V_{ob}\left(t_0\right) e^{-\left\{\rho_{ob}\left(t-t_0\right)\right\}} + \ \ell_{ob}\ {\rho_{ob}}^{-1} \tag{45}$$

From (40):

$$\|\boldsymbol{e}_{ob}\|^2 \leq 2V_{ob}\left(t_0\right) e^{-\left\{\rho_{ob}\left(t-t_0\right)\right\}} + 2\ \ell_{ob}\ {\rho_{ob}}^{-1} \tag{46}$$

Based on Minkowski's inequality:

$$\|\boldsymbol{e_{ob}}\| \leq \sqrt{2V_{ob}\left(t_0\right)} e^{-\frac{\rho_{ob}}{2}(t-t_0)} + \sqrt{2\ell_{ob}\rho_{ob}^{-1}} \tag{47}$$

Thus, based on Corless and Leitmann (1993), it is clear from (47) that the vector of the estimation errors $\|\boldsymbol{e_{ob}}\|$ reaches a defined region $g_0\left(\tau_0\right)$ in uniformly exponential convergence, such that

$$g_0\left(\tau_0\right) := \left\{\|\boldsymbol{e_{ob}}\| \leq \bar{\tau}_0 := \sqrt{2\ell_{ob}\rho_{ob}^{-1}}\right\} \tag{48}$$

## Appendix B. Proof of Theorem 3.1

By using the proposed HAC-BLFs for each $SS_i$ in (14), a quadratic function is defined for $SS_1$, as Tee et al. (2009)

$$\bar{V}_1 = \frac{1}{2} \log\left(\frac{o_1^2}{Q_1}\right) + \frac{1}{2}\tilde{\theta}_1^2 \tag{49}$$

Given that the initial estimated state $\hat{x}_1\left(t_0\right)$ and the reference trajectory $x_{1d}\left(t_0\right)$ can be specified, they are selected to satisfy $\bar{e}_1\left(t_0\right) = \hat{x}_1\left(t_0\right) - x_{1d}\left(t_0\right) < o_1$. By inserting \ eqref {15} and \ eqref {3000}, and using the condition $\bar{e}_0 = 0$, it follows that:

$$\dot{\bar{V}}_1 = \frac{\bar{e}_1}{Q_1}\bar{e}_2 - \frac{1}{2}\gamma_1 \frac{\bar{e}_1}{Q_1}\bar{e}_1 - \epsilon_1\tilde{\theta}_1(\frac{\bar{e}_1}{Q_1})^2 + \epsilon_1\tilde{\theta}_1(\frac{\bar{e}_1}{Q_1})^2 - \kappa_1\tilde{\theta}_1^2 \tag{50}$$

**Lemma B.1.** *For any positive constant $o_i$, and given that $Q_i = o_i^2 - \bar{e}_i^2$, the following inequality holds for all $\bar{e}_i$ satisfying $|\bar{e}_i| < o_i$ (Ren et al., 2010):*

$$\log\left(\frac{o_i^2}{Q_i}\right) < \frac{\bar{e}_i^2}{Q_i} \tag{51}$$

Using Lemma B.1:

$$\dot{\bar{V}}_1 \leq \frac{\bar{e}_1}{Q_1}\bar{e}_2 - \frac{1}{2}\gamma_1 \log\left(\frac{o_1^2}{Q_1}\right) - \kappa_1\tilde{\theta}_1^2 \tag{52}$$

From (49) and (20):

$$\dot{\bar{V}}_1 \leq \bar{s}_1 - \bar{\rho}_1\bar{V}_1, \quad \bar{\rho}_1 = \min[\gamma_1,\ 2\kappa_1], \tag{53}$$

Similarly, quadratic functions are defined for each $SS_i$ for $i = 2, \ldots, n-1$, as

$$\bar{V}_i = \frac{1}{2} \log\left(\frac{o_i^2}{Q_i}\right) + \frac{1}{2}\tilde{\theta}_i^2 \tag{54}$$

By inserting $\bar{e}_i(t_0) < o_i$ to (15) and (17):

$$\dot{\bar{V}}_i = \frac{\bar{e}_i}{Q_i}\bar{e}_{i+1}\dot{\bar{e}}_i + \tilde{\theta}_i\dot{\tilde{\theta}}_i = \frac{\bar{e}_i}{Q_i}\bar{e}_{i+1} - \frac{1}{2}\gamma_i\frac{\bar{e}_i}{Q_i}\bar{e}_i - \epsilon_i\tilde{\theta}_i(\frac{\bar{e}_i}{Q_i}\bar{e}_{i+1})^2 \\ - \ \frac{\bar{e}_i}{Q_i}\frac{Q_i}{Q_{i-1}}\bar{e}_{i-1} - \kappa_i\tilde{\theta}_i^2 + \epsilon_i\tilde{\theta}_i(\frac{\bar{e}_i}{Q_i}\bar{e}_{i+1})^2 \tag{55}$$

Thus,

$$\dot{\bar{V}}_i = \frac{\bar{e}_i}{Q_i}\bar{e}_{i+1} - \frac{1}{2}\gamma_i\frac{\bar{e}_i^2}{Q_i} - \frac{\bar{e}_i}{Q_{i-1}}\bar{e}_{i-1} - \kappa_i\tilde{\theta}_i^2 \tag{56}$$

Using Lemma B.1, and (20),

$$\dot{\bar{V}}_i = \bar{s}_i - \frac{1}{2}\gamma_i \log\left(\frac{o_i^2}{Q_i}\right) - \bar{s}_{i-1} - \kappa_i\tilde{\theta}_i^2 \tag{57}$$

Therefore,

$$\dot{\bar{V}}_i \leq \bar{s}_i - \bar{\rho}_i\bar{V}_i - \bar{s}_{i-1} + \bar{\ell}_i, \quad \bar{\rho}_i = \min[\gamma_i,\ 2\kappa_i] \tag{58}$$

Similarly, another quadratic function is defined for $SS_n$, as

$$\bar{V}_n = \frac{1}{2} \log\left(\frac{o_n^2}{Q_n}\right) + \frac{1}{2}\tilde{\theta}_n^2 \tag{59}$$

By selecting $\bar{e}_n(t_0) < o_n$ and inserting (15) and (17):

$$\dot{\bar{V}}_n = \frac{\bar{e}_n}{Q_n}\dot{\bar{e}}_n + \tilde{\theta}_n\dot{\tilde{\theta}}_i = \frac{\bar{e}_n}{Q_n}\Delta_u - \frac{1}{2}\gamma_n\frac{\bar{e}_n}{Q_n}\bar{e}_n - \epsilon_n\tilde{\theta}_n(\frac{\bar{e}_n}{Q_n})^2 \\ - \ \frac{\bar{e}_n}{Q_n}\frac{Q_n}{Q_{n-1}}\bar{e}_{n-1} - \kappa_n\tilde{\theta}_n^2 + \epsilon_n\tilde{\theta}_n(\frac{\bar{e}_n}{Q_n})^2 \tag{60}$$

By defining an unknown positive constant as $\alpha_{max} = |\Delta_u|$:

$$\dot{\bar{V}}_n \leq \frac{|\bar{e}_n|}{Q_n}\alpha_{max} - \frac{1}{2}\gamma_n\frac{\bar{e}_n^2}{Q_n} - \frac{\bar{e}_n}{Q_{n-1}}\bar{e}_{n-1} - \kappa_n\tilde{\theta}_n^2 \tag{61}$$

Based on Young's inequality and arbitrary positive constant $\upsilon_n$, and knowing that $Q_n$ is a positive notation:

$$\dot{\bar{V}}_n \leq \frac{\upsilon_n}{2}\frac{\bar{e}_n^2}{Q_n} + \frac{\alpha_{max}^2}{2\upsilon_n Q_n} - \frac{1}{2}\gamma_n\frac{\bar{e}_n^2}{Q_n} - \frac{\bar{e}_n}{Q_{n-1}}\bar{e}_{n-1} - \kappa_n\tilde{\theta}_n^2 \tag{62}$$

From (20) and (59), for any $\gamma_n > \upsilon_n$:

$$\dot{\bar{V}}_n \leq -\bar{s}_{n-1} - \bar{\rho}_n\bar{V}_n + \frac{\alpha_{max}^2}{2\upsilon_n Q_n}, \\ \bar{\rho}_n = \min[\gamma_n - \upsilon_n,\ 2\kappa_n] \tag{63}$$

Now, a quadratic function is defined for all proposed HAC-BLFs employed by the reference system (12), as

$$\bar{V}_{cont} = \sum_{i=1}^{n}\bar{V}_i = \sum_{i=1}^{n}\frac{1}{2}\log\left(\frac{o_i^2}{Q_i}\right) + \frac{1}{2}\tilde{\theta}_i^2 \tag{64}$$

The derivative of (64), based on (53), (58), and (63), can be expressed as

$$\dot{\bar{V}}_{cont} \leq \sum_{i=1}^{n-1}\left[-\bar{s}_i + \bar{s}_i\right] - \bar{\rho}_{cont}\bar{V}_{cont} + \bar{\ell} \tag{65}$$

where

$$\bar{\rho}_{cont} = \min[\bar{\rho}_1, \dots, \bar{\rho}_n] = \min[\min_{i=1}^{n}\{\gamma_i, 2\kappa_i\}], \qquad \bar{\ell} = \frac{\alpha_{max}^2}{2\upsilon_n Q_n} \tag{66}$$

As observed, similar to (41), term $e_{i+1}$ for $SS_i$ is treated as an external input that causes $\bar{s}_i$ to appear in the $i$th SS. Hence, by using the proposed control (14), instability terms $\bar{s}_i$ in each SS, provided in (20), will be canceled out for the entire system. Thus:

$$\dot{\bar{V}}_{cont} \leq -\bar{\rho}_{cont}\bar{V}_{cont} + \bar{\ell} \tag{67}$$

(67) can be solved as follows:

$$\bar{V}_{cont} \leq \bar{V}_{cont}\left(t_0\right) e^{-\left\{\bar{\rho}_{cont}\left(t-t_0\right)\right\}} + \bar{\ell}\int_{t_0}^{t} e^{\left\{-\bar{\rho}_{cont}(t-T)\right\}} dT \tag{68}$$

Because $e^{-\bar{\rho}_{cont}(t-T)}$ is always decreasing,

$$\bar{V}_{cont} \leq \bar{V}_{cont}\left(t_0\right) e^{-\left\{\bar{\rho}_{cont}\left(t-t_0\right)\right\}} + \bar{\ell}\ \bar{\rho}_{cont}^{-1} \tag{69}$$

From (64):

$$\sum_{i=1}^{n} \log\left(\frac{o_i^2}{Q_i}\right) \leq 2\bar{V}_{cont}\left(t_0\right) e^{-\left\{\bar{\rho}_{cont}\left(t-t_0\right)\right\}} + 2\ \bar{\ell}\ \bar{\rho}_{cont}^{-1} \tag{70}$$

Thus, $\sum_{i=1}^{n}\log\left(\frac{o_i^2}{Q_i}\right)$ reaches a defined region $g_1\left(\tau_1\right)$ in uniformly exponential convergence, such that

$$g_1\left(\tau_1\right) := \left\{\sum_{i=1}^{n}\log\left(\frac{o_i^2}{Q_i}\right) \leq \bar{\tau}_1 := 2\ \bar{\ell}\ \bar{\rho}_{cont}^{-1}\right\} \tag{71}$$

## Appendix C. Proof of Theorem 4.1

A quadratic function is defined for the uncertain system (2) employing HAEs and HAC-BLFs as

$$V_{all} = V_{ob} + \bar{V}_{cont} = \sum_{i=1}^{n}\frac{1}{2}\left[e_i^2 + \tilde{\Psi}_i^2 + \log\left(\frac{o_i^2}{Q_i}\right) + \tilde{\theta}_i^2\right] \tag{72}$$

Based on (41) and (65):

$$\dot{V}_{all} \leq -\rho_{ob}V_{ob} + \ell_{ob} - \bar{\rho}_{cont}\bar{V}_{cont} + \bar{\ell} \tag{73}$$

Straightforwardly:

$$\dot{V}_{all} \leq -\rho_{all}V_{all} + \ell_{all} \tag{74}$$

where $\rho_{all} = \min[\rho_{ob}, \bar{\rho}_{cont}]$ and $\ell_{all} = \ell_{ob} + \bar{\ell}$. (67) can be solved as follows:

$$V_{all} \leq V_{all}\left(t_0\right) e^{-\left\{\rho_{all}\left(t-t_0\right)\right\}} + \ell_{all}\int_{t_0}^{t} e^{\left\{-\rho_{all}(t-T)\right\}} dT \tag{75}$$

From (74):

$$\sum_{i=1}^{n} e_i^2 + \log\left(\frac{o_i^2}{Q_i}\right) \leq 2V_{all}\left(t_0\right) e^{-\rho_{all}\left(t-t_0\right)} + 2\ell_{all}\rho_{all}^{-1} \tag{76}$$

Therefore, $\sum_{i=1}^{n} e_i^2 + \log\left(\frac{o_i^2}{Q_i}\right)$ with initial time $t_0$ converging uniformly and exponentially within a specific region $g_2\left(\tau_2\right)$ centered at zero such that

$$g_2\left(\tau_2\right) := \left\{\sum_{i=1}^{n} e_i^2 + \log\left(\frac{o_i^2}{Q_i}\right) \leq \bar{\tau}_2 := \sqrt{2\ \ell_{all}\ \rho_{all}{}^{-1}}\right\} \tag{77}$$

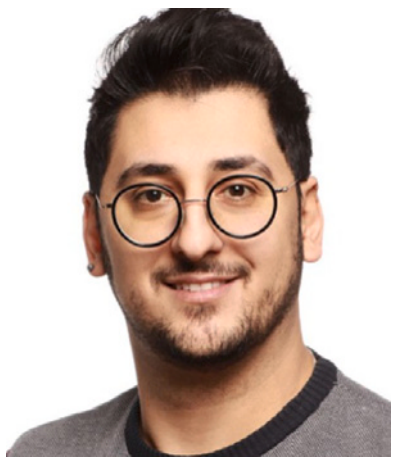

**Mehdi Heydari Shahna** earned a B.Sc. in electrical engineering from Razi University, Kermanshah, Iran, in 2015 and an M.Sc. in control engineering at Shahid Beheshti University, Tehran, Iran, in 2018. Since December 2022, he has been pursuing his doctoral degree in automation technology and mechanical engineering at Tampere University, Tampere, Finland. His research interests encompass robust control, robotics, fault-tolerant algorithms, and system stability, as well as robot learning.

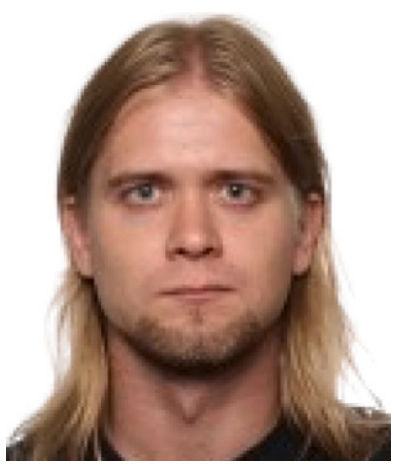

**Jukka-Pekka Humaloja** received a Ph.D. in 2019 and worked as a postdoctoral research fellow until 2021 in the Systems Theory Research Group, Tampere University, Finland. During 2021–2023, he was a postdoctoral fellow in the Distributed Parameter System Lab at the University of Alberta, Canada. He is currently a postdoctoral researcher at the Technical University of Crete, Greece. His research interests include nonlinear control, control of distributed parameter systems, and advanced control strategies for complex systems.

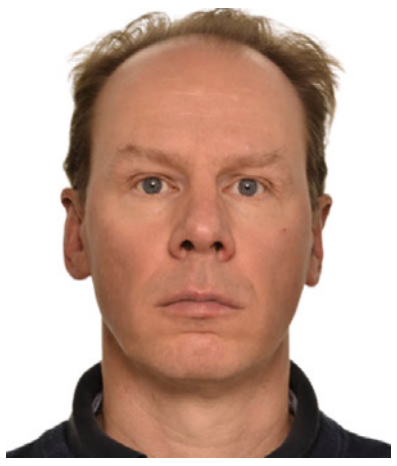

**Jouni Mattila** received an M.Sc. and Ph.D. in automation engineering from Tampere University of Technology, Tampere, Finland, in 1995 and 2000, respectively. He is currently a professor of machine automation in the Automation Technology and Mechanical Engineering Unit at Tampere University. His research interests include machine automation, nonlinear-model-based control of robotic manipulators, and energy-efficient control of heavy-duty mobile manipulators.